\documentclass[useAMS,usenatbib,usegraphicx]{mn2e}
\usepackage{journals}
\usepackage{times}
\usepackage{color}
\definecolor{green}{rgb}{0,0.5,0}
\definecolor{grey}{rgb}{0.4,0.5,0.7}

\renewcommand{\d}{{\rm d}}
\title[Modelling Anisotropy and Mass Profiles]
{\emph{MAMPOSSt}: \emph{M}odelling \emph{A}nisotropy and \emph{M}ass \emph{P}rofiles
  of \emph{O}bserved \emph{S}pherical \emph{S}ys\emph{t}ems. I. Gaussian 3D
  velocities} 
\author[Mamon et al.]
{Gary A.
    Mamon$^{1,2}$,
Andrea Biviano$^3$
and
Gwena\"el Bou\'e$^{4,5,6,1}$\\
$^1$ Institut
d'Astrophysique de
    Paris (UMR 7095: CNRS \& UPMC), 98 bis Bd Arago, F--75014 Paris, France,
    {e-mail: gam@iap.fr},
    \\
$^2$ Astrophysics \& BIPAC, Department of Physics, University of Oxford, Oxford OX1 3RH, UK\\
$^3$ INAF/Osservatorio Astronomico di Trieste, via G.B. Tiepolo 11, 34143
    Trieste, Italy\\
$^4$ Astronomie et Syst\`emes Dynamiques, IMCCE-CNRS UMR8028,
             Observatoire de Paris, UPMC, 77 Av. Denfert-Rochereau,
             75014 Paris, France\\
$^5$ Centro de Astrof\'{\i}sica da Universidade do Porto, Rua das Estrelas 
4150-762 Porto, Portugal\\
$^6$ Department of Astronomy and Astrophysics, University of
             Chicago, 5640 South Ellis Avenue, Chicago, IL 60637, USA
}

\date{Accepted 2012 Dec 05. Received 2012 Dec 05; in original form 2012 Jul 20}
\pubyear{2012}
\begin{document}
\maketitle

\begin{abstract}
Mass modelling of spherical systems through internal kinematics is
hampered by the mass / velocity anisotropy degeneracy inherent in the
Jeans equation, as well as the lack of techniques that are both fast and
adaptable to realistic systems.
A new fast method, called MAMPOSSt, is developed and
thoroughly tested. MAMPOSSt performs a maximum likelihood fit of the
distribution of observed tracers in projected phase space (projected
radius and line-of-sight velocity). As in other methods, MAMPOSSt
assumes a shape for the gravitational potential (or equivalently the
total mass profile).  However, instead of postulating a shape for the
distribution function in terms of energy and angular momentum, or
supposing Gaussian line-of-sight velocity distributions, MAMPOSSt
assumes a velocity anisotropy profile and a shape for the
three-dimensional velocity distribution.  The formalism is presented
for the case of a Gaussian 3D velocity distribution. 
In contrast to most methods based on moments,
MAMPOSSt requires no binning, differentiation, nor extrapolation of the observables.
Tests on cluster-mass haloes from $\Lambda$CDM dissipationless
cosmological simulations indicate that, with 500 tracers, MAMPOSSt is able to
jointly recover the virial radius, tracer scale radius, dark matter scale
radius and outer or constant velocity anisotropy with small bias ($<$10\% on scale radii and
$<$2\% on the two other quantities) and
inefficiencies 
of 10\%, 27\%, 48\% and 20\%, respectively.
MAMPOSSt does not perform better when some parameters are frozen, and even particularly
worse when the virial radius is set to its true value, which appears to be
the consequence of halo triaxiality.
The accuracy of MAMPOSSt depends weakly on the adopted interloper removal
scheme, including an efficient iterative Bayesian scheme that we introduce here, which
can directly obtain the virial radius with as good precision as MAMPOSSt.
Additional tests are made on the number of tracers, the stacking of haloes, 
the chosen aperture, and the
density and velocity anisotropy models.
Our tests show that MAMPOSSt with Gaussian 3D velocities
is very competitive with other methods that are either currently
restricted to constant velocity anisotropy or 3 orders of magnitude slower.
These tests suggest that MAMPOSSt can
be a very powerful and rapid method for the mass and anisotropy
modeling of systems such as clusters and groups of galaxies, elliptical
and dwarf spheroidal galaxies.

\end{abstract}


\section{Introduction}

The determination of mass profiles is one of the fundamental issues of
astronomy. Subtracting the mass density profile of the visible component, one deduces
the dark matter (hereafter, DM) density profile, which can be confronted to the predictions from
cosmological $N$-body 
simulations. This is especially relevant given the differences
between the total NFW \citep*{NFW96} or better Einasto \citep{Navarro+04}
density profiles derived in dissipationless simulations of a single dark
matter component on one hand, and the $1/r^2$ density profiles found for
the DM in hydrodynamical cosmological simulations \citep{GKKN04}.
Moreover, the knowledge of the total density profile serves as a fundamental
reference, relative to which one can scale various
astronomical tracers such as the mass density profiles of the stellar, gas
and dust components, as well as the luminosity in different wavebands.
These studies can be performed as a function of system mass and other
attributes such as galaxy colour (e.g., \citealp{WM13} and references therein).

Mass profiles can be derived from internal motions, or alternatively from
X-ray or lensing observations. This paper focuses on mass profiles from
internal kinematics. In this class of mass modeling, one has to deal with a
degeneracy between the unknown radial profiles of total mass and of the
velocity 
anisotropy (hereafter `anisotropy')
\begin{equation}
\beta(r) = 1 - {\sigma_\theta^2(r) + \sigma_\phi^2(r)  \over
  2\,\sigma_r^2(r)} 
\label{beta}
\end{equation}
(note that, in spherical symmetry, one must have $\sigma_\phi =
\sigma_\theta$).
While radial outer orbits are expected for structures in an expanding
universe (e.g. \citealp{AG08} for dark matter particles and
\citealp{Ludlow+09} for subhaloes), the dissipative nature of the gas dynamics
is expected to produce tangential orbits in the inner regions of systems
formed from gas-rich mergers or collapse. 
Therefore,  lifting the \emph{Mass - Anisotropy Degeneracy}
can provide useful constraints on the formation of
the structure under study. 

A common method to extract the mass profile is to assume that the 
line-of-sight (hereafter, LOS) velocity distribution, at given projected radius, is Gaussian
\citep{Strigari+08,Battaglia+08,Wolf+10}. 
These methods perform adequately on the mass profile, but provide weak
constraints on the anisotropy \citep{Walker+09}.
\cite{Merritt87} pointed out that anisotropic models have non-Gaussian
LOS velocity distributions. Therefore, the observed kurtosis of the
distribution of LOS velocities serves as a powerful constraint to
the anisotropy \citep*{Gerhard93,vdMF93,ZFG93}. 
In fact, 
if one assumes that the anisotropy is constant throughout the
system, the fourth order Jeans equation can be used to express
the LOS velocity kurtosis as an integral
of the tracer density, anisotropy and total mass profiles
\citep{Lokas02}. Moreover,
\cite{RF12} were able to generalize the expression
  for the LOS velocity kurtosis for radially varying anisotropy in the
  framework of separable augmented density and 4th order anisotropy equal to
  the standard anisotropy (the latter appears to be an excellent
  approximation for $\Lambda$CDM haloes, see Fig.~10 of \citealp{Wojtak+08}).
One can then perform a joint fit of the observed LOS velocity dispersion and
kurtosis profiles. This was found to (partially) lift
the mass-anisotropy degeneracy when applied to dwarf spheroidal galaxies
\citep{Lokas02} and the Coma cluster, \citep{LM03}:
the joint constraint of LOS velocity dispersion and
kurtosis profiles allows the  estimation of both the mass  profile
(i.e., normalization and concentration) and the
anisotropy of the cluster, contrary to the case when the LOS
kurtosis profile is  ignored. 

An interesting route is to perform non-parametric inversions of the data
assuming either the mass profile to obtain the anisotropy profile (anisotropy
inversion, pioneered by \citealp{BM82}) or the anisotropy profile to obtain
the mass profile (mass inversion, independently developed by  \citealp{MB10}
and \citealp{Wolf+10}). These inversion methods are
powerful in that they are non-parametric, but they suffer from their
requiring the user to
bin the data, smooth it, and extrapolate it beyond the range of data.

Hence, one would like to go one step further and constrain the full information
contained in the observed projected phase space (projected radii and
LOS velocities, hereafter, PPS) 
of LOS velocities as a
function of projected radii. In other words, rather than using the 0th, 2nd
and possibly 4th moments of the LOS velocity distribution, we wish
to use the full set of even moments.

The traditional way to analyze the distribution of particles in PPS is to
assume a form for the six-dimensional distribution function (DF) in terms of
energy ($E$) and angular momentum ($J$) and fit the triple integral
of equation~(\ref{gDM}) below, using this DF for $f$,
to the 
distribution of particles in PPS.
The worry is that we have no good a priori knowledge
of the shape of the DF, $f(E,J)$. One clever idea is to throw orbits in a gravitational
potential,
since each orbit is a Dirac delta function in energy and angular
momentum. One then seeks a linear combination of these orbits, with positive
coefficients, to match the data. This orbit model
\citep{Schwarzschild79,RT84,ST96} is very powerful (and can handle non-spherical
gravitational potentials), but too
slow to obtain meaningful errors on the parameters.
A similar, and in principle faster, technique is to assume that the DF is the
linear combination (again with positive coefficients) of
elementary DFs \citep{Dejonghe89,MS93,GJSB98}, but only one such study has been made
\citep{KSGB00}, and it is not clear whether the elementary DFs, although
numerous, constitute a basis set.

An important step forward has been performed by \cite{Wojtak+08}, who
analyzed the haloes in $\Lambda$CDM cosmological simulations, to show that the
DF can be approximated to be separable in energy and
angular momentum, with a simple analytical approximation for the angular
momentum term. In a sequel, \cite{WLMG09} have shown that it is feasible
to fit the distribution of particles in PPS with
equation~(\ref{gDM}), using the approximation of the DF
found by \nobreak{\cite{Wojtak+08}.}

However, it is not yet clear whether self-gravitating quasi-spherical
astrophysical systems have the DF of $\Lambda$CDM haloes:
In particular, if the dynamical evolution of these systems is influenced by
the dissipation of their gaseous component, the DF may not be separable in terms
of energy and angular momentum. 
Dissipation is not expected to
affect much the internal kinematics of large systems such as galaxy clusters\footnote{However,
the joint X-ray and lensing analysis of a cluster by \cite{Newman+09}
reveals a shallower inner density profile than NFW, suggesting that
dissipation is also important in clusters.}, but
is expected to be increasingly important in smaller systems such as galaxy
groups, and especially 
galaxies themselves. 
For this reason, it is useful to consider a mass-modeling method that is
independent of the dependence of the DF on energy and
angular momentum. 

In this work, we present an alternative method, in which we
fit the distribution of particles in PPS making assumptions
on the radial profiles of mass and anisotropy as well as the radial
variations of the distribution of space-velocities.
We call this method  \emph{Modelling of Anisotropy and Mass
  Profiles of Observed Spherical Systems}, or \emph{MAMPOSSt} for
short.\footnote{MAMPOSSt should evoke the mass analog of a lamppost, and
  \emph{mamposter\'{\i}a} in Spanish means masonry, hence the building
  blocks of structures.}
The MAMPOSSt method is described in Sect.~\ref{genmeth}, its Gaussian
approximation is described in Sect.~\ref{gaussian}.  Tests on haloes
derived from a cosmological $N$ body simulation are presented in
Sect.~\ref{tests}. A discussion follows in Sect.~\ref{discus}.

\section{Method}

\subsection{General method}
\label{genmeth}
The observed \emph{tracer population} of a spherical system has a
DF
\begin{equation}
f({\bmath r},{\bmath v}) = \nu(r)\,f_v({\bmath v}|r) \ ,
\label{fsep}
\end{equation}
where $\nu(r)$ is the tracer number density profile.

MAMPOSSt fits the distribution of objects in PPS
(projected radius $R$ and LOS velocity $v_z$),
assuming parametrized forms for
 \begin{enumerate}
\item the
gravitational 
potential (or equivalently a total mass density profile, through the Poisson
equation),
\item  the anisotropy profile (eq.~[\ref{beta}]) 
\item the distribution of 3D velocities, $f_v({\bmath v}|r)$.
\end{enumerate}

Consider a point P at distance $r$ from the centre, O, of the spherical system,
with projected radius $R \leq r$ and consider the spherical
coordinates where the unit
vectors $\bmath e_r$ and $\bmath e_\theta$ are in the plane containing OP and the
LOS, while 
$\bmath e_\phi$ is perpendicular to this plane.
Consider also the cylindrical coordinate system $(v_z,v_\perp,v_\phi)$, where
${\bmath e_z}$ is the axis along the LOS and ${\bmath e_\perp}$ is the axis
perpendicular to the LOS, but in the plane containing O and P and the LOS.
The Jacobian of the transformation from the spherical coordinate system to
the new one is unity, hence one can write
\begin{eqnarray}
f_v(v_z,v_\perp,v_\phi|\{r,R\}) &\equiv &
\left ({{\rm d}^3 N \over \d v_z\, \d v_\perp\, \d v_\phi}\right)_{r,R} \nonumber \\
&=&\left ( {{\rm d}^3 N \over \d v_r \,\d v_\theta \,\d v_\phi }
\right)_{r,R} \nonumber \\ 
&\equiv& 
f_v(v_r,v_\theta,v_\phi|\{r,R\}) \ . \nonumber
\end{eqnarray}

The distribution of LOS velocities at P is then obtained by
integrating velocities over the two perpendicular axes (dropping $\{r,R\}$
from $f_v$ 
for clarity):
\begin{equation}
h(v_z|R,r) \equiv \left ({\d N\over \d v_z}\right)_{r,R} 
=
\int_{-\infty}^{+\infty} 
\d v_\perp 
\int_{-\infty}^{+\infty} 
f_v(v_z,v_\perp,v_\phi) \,\d v_\phi \,.
\label{dNdvz}
\end{equation}
Note that dynamical systems have maximum velocities set by the escape
velocity, $\sqrt{-2\,\Phi(r)}$ (where $\Phi(r)$ is the gravitational
potential), on one hand, and by the maximum allowed (observable) absolute LOS
velocity on the other hand.
In what follows, we will neglect both limits, unless explicitly mentioned otherwise.

The surface density of observed 
objects (\emph{the tracer}) in PPS is then obtained by
integrating along the LOS
\begin{eqnarray}
g(R,v_z) 
&\!\!\!\!=\!\!\!\!& 
\Sigma(R)\,\left \langle h(v_z|R,r) 
\right\rangle_{\rm LOS} \nonumber \\
&\!\!\!\!=\!\!\!\!& 
2\int_R^\infty {r\,\nu(r)\over
  \sqrt{r^2-R^2}}\,h(v_z|R,r)\,\d r
\label{g1}\\
&\!\!\!\!=\!\!\!\!& 
2\int_R^\infty
\!\!\!\!
{r\,\d r\over \sqrt{r^2-R^2}}
\int_{-\infty}^{+\infty} 
\!\!\!\!\!\!\!
\d v_\perp 
\!\!
\int_{-\infty}^{+\infty} 
\!\!\!\!\!\!\!
f(r,v_z,v_\perp,v_\phi) \,\d v_\phi \,.
\label{gDM}
\end{eqnarray} 
where 
\begin{equation}
\Sigma(R) = \int_{-\infty}^{+\infty} \nu(r)\,\d z = 2\,\int_R^\infty {r\,\nu(r)
  \,\d r  \over \sqrt{r^2-R^2}}
\label{Sigma}
\end{equation}
is the tracer surface density at projected radius $R$.
Equation~(\ref{gDM}) is equivalent to equation~(2) of \cite{DM92}.

If the tracer number density
profile $\nu(r)$, appearing in equation~(\ref{g1}), is not known and if the incompleteness of the data is
independent of the projected radius, then one can estimate 
$\nu(r)$  by Abel inversion of $\Sigma(R)$ of equation~(\ref{Sigma}): 
\begin{equation}
\nu(r) = -{1\over \pi} \,\int_r^\infty {\d \Sigma\over \d R}\, {\d R\over
  \sqrt{R^2-r^2}} 
  \ .
\label{rho}
\end{equation}
But this is not necessary, as we shall see below.

In MAMPOSSt,
rather than replace the velocities by energy and angular momentum and
numerically solve 
the triple integral of equation~(\ref{gDM}) (as first
proposed by \citeauthor{DM92}, see also \citealp{WLMG09}), we analytically
derive $h(v_z|R,r)$ from equation~(\ref{dNdvz}) 
for known 3D 
velocity distributions.
With the analytical form of $h(v_z|R,r)$, 
equation~(\ref{g1}) provides the surface density distribution of tracers
in PPS through a single integral.
Note, however, that another single integral is required because the expression for 
$h(v_z|R,r)$ 
will involve 
$\sigma_r(r)$
(see eqs.~[\ref{hgauss}] and [\ref{sigmaz}], below, for
the Gaussian case), which is obtained 
by solving the spherical Jeans
equation
\begin{equation}
{\d \left (\nu \sigma_r^2\right) \over \d r} + 2\,\beta\,{\nu \sigma_r^2\over r} = -\nu (r)\,{G
  M(r)\over r^2}
\label{jeans}
\end{equation} 
where $\beta$ is the anisotropy of (eq.~[\ref{beta}])  for our given choices of total mass and anisotropy profiles.
We thus need to insert the solution \citep{vanderMarel94,ML05b}
\begin{equation}
\sigma_r^2(r) = {1\over \nu(r)}\,\int_r^\infty \exp\left [2\,\int_r^s \beta(t) {\d t\over
    t}\right]\,\nu(s) {G M(s) \over s^2}\,\d s \ ,
\label{sigmar2}
\end{equation}
in the
expression for $h(v_z|R,r)$ (eq.~[\ref{dNdvz}]) 
to derive $g(R,v_z)$, via equation~(\ref{g1}), where $\beta(t)$ is given, while
$\nu(r)$ is obtained with equation~(\ref{rho}). 
In equation~(\ref{sigmar2}), $M(s) = (s^2/G) \,\d\Phi/\d s$ is the radial profile of the 
\emph{total} mass
(this is the only instance where the gravitational potential enters
MAMPOSSt).
For a given choice of parameters,
the single integral of equation~(\ref{g1}) must be evaluated for
every data point $(R,v_z)$, whereas the other integral (eq.~[\ref{sigmar2}]) for
$\sigma_r(r)$ need only be evaluated once, on an
adequate grid of $r$.  

Note that for projected radii extending from $R_{\rm min}$ to $R_{\rm max}$
and absolute LOS velocities extending from 0 to a maximum velocity, which for
projected radius $R$ is 
theoretically equal to $v_{\rm esc }(R) = \sqrt{-2\Phi(R)}$, and in practice is 
possibly specified by a cut of obvious velocity interlopers, $v_{\rm
  cut}(R)$, one can 
write 
\begin{eqnarray}
\int_{R_{\rm min}}^{R_{\rm max}}2\pi \,R\,\d R \int_{-v_{\rm cut}R)}^{v_{\rm cut}(R)}
\!\!\!\!  g(R,v_z)\,\d v_z & 
\!\!\!\! 
= 
\!\!\!\! 
& 2\pi\,\int_{R_{\rm min}}^{R_{\rm max}} 
\!\!\!\!
R\,\Sigma(R)\,\d R
 \nonumber \\
& 
\!\!\!\! 
= 
\!\!\!\! 
& \Delta N_{\rm p} \ ,
\label{gint}
\end{eqnarray}
where we used equation~(\ref{g1}) for $g(R,v_z)$, assumed that $h(v_z|R,r)$
is normalised, reversed the order of the integrals in $r$ and $v_z$,
and where $N_{\rm p}(R)$ is the predicted number of objects within 
  projected radius $R$, while 
$\Delta N_{\rm p} = N_{\rm p}(R_{\rm max})\!-\!N_{\rm p}(R_{\rm min})$.
Equation~(\ref{gint}) then implies that the probability density of observing an object at
position $(R,v_z)$ of PPS is
\begin{eqnarray}
q(R,\!v_z)\!\!\!&\!\!\!\!\!\!\!=\!\!\!\!\!\!\!&\!\!\!{2 \pi\,R\,g(R,v_z) \over \Delta
  N_{\rm p}} \nonumber \\ 
&\!\!\!\!\!\!\!=\!\!\!\!\!\!\!&\!\!\!{4\pi\,R\over \Delta N_{\rm p}}\,\int_R^\infty
{r \,\nu(r)\over 
  \sqrt{r^2-R^2}}\,h(v_z|R,r)\,\d r 
\label{qdef}\\
&\!\!\!\!\!\!\!=\!\!\!\!\!\!\!&\!\!\!{R^2\over \Delta \widetilde N_{\rm p} r_\nu^3}
\!\!\int_0^\infty
\!\!\!\!\!\!
\cosh u\,\widetilde \nu\!\left({R\over r_\nu} \cosh u\right)\!h(v_z|R,R\cosh u)
\,{\rm d}u,\nonumber \\ 
\label{qdef2}
\end{eqnarray} 
where equation~(\ref{qdef}) arises from equation~(\ref{g1}), 
while equation~(\ref{qdef2}) is obtained by writing $r = R\cosh u$. 
Here, $N_{\rm p}(R)$ is the number of tracers in a cylinder of projected
radius $R$, 
the terms
$\widetilde \nu$ and $\widetilde N_{\rm p}$ are given by
\begin{eqnarray}
\nu(r) = {N\left(r_\nu \right) \over 4 \pi r_\nu^3}\,
\widetilde \nu\left ({r\over r_\nu}\right) \ ,
\\
N_{\rm p}(R) = N\left (r_\nu\right)\,\widetilde N_{\rm p} \left
({R\over r_\nu}\right) \ ,
\end{eqnarray}
where $N(r)$ is the cumulative tracer number density profile,
while
$r_\nu$ is the characteristic radius of the tracer.
One easily verifies that $\int\!\!\int q(R,v_z)\, \d R\,\d v_z=1$.
The values of $R_{\rm min}$ and $R_{\rm max}$ appearing in $\Delta N_{\rm p}$
(eq.~[\ref{gint}]) can be hard limits, or alternatively the respective
minimum and
maximum projected radii of the observed tracers if no hard limits are specified. 

We fit the parameters (mass scale or concentration and possibly  normalization,
anisotropy level or radius, as well as 
the tracer scale
$r_\nu$
-- if not previously known)
that enter the determination of  $g(R,v_z)$ to the observed 
surface density, using maximum
likelihood estimation (MLE), i.e. by minimizing
\begin{equation}
-\ln {\cal L} = - \sum_{i=1}^n \ln q(R_i,v_{z,i}|\bmath{\btheta}) \ ,
\label{lnL}
\end{equation}
for the $N$-parameter vector $\bmath{\btheta}$, where $n$ is the number
of data points, with $q$ given by
equation~(\ref{qdef}).

Writing $\bmath{\theta} = \{r_\nu, \bmath{\eta}\}$, where
$\bmath{\eta}$ is the vector of the $N\!-\!1$ parameters other than $r_\nu$,
one has
\begin{equation}
q(R,v_z|r_\nu,\bmath{\eta}) = p_0(R|r_\nu) \times
p(v_z|R,r_\nu,\bmath{\eta})
\ ,
\label{sumsum}
\end{equation}
where
\begin{equation}
p_0(R|r_\nu) = {2 \pi \,R\,\Sigma(R|r_\nu) 
\over N_p(R_{\rm max}|r_\nu)  -  N_p(R_{\rm min}|r_\nu) }
\ 
\label{pofR1}
\end{equation}
and
\begin{equation}
p(v_z|R,\bmath{\eta}) \equiv \left \langle 
h(v_z|R,r)\right\rangle_{\rm LOS}
=
{g(R,v_z|\bmath{\btheta})\over
  \Sigma(R|r_\nu)} \ .
\label{pofvzgivenR1}
\end{equation}
Combining the last equality of equation~(\ref{pofvzgivenR1})
with equation~(\ref{gDM}), integrating over LOS velocities,
reversing the order of the two outer integrals
of the resulting
quadruple integral, and using equation~(\ref{Sigma})  yields $\int p(v_z|R)\,\d v_z=1$.   
So, if the scale of the
tracer distribution is already known, then, according to
equations~(\ref{lnL}) and (\ref{sumsum}),  maximizing the
likelihood amounts to minimizing
\begin{equation}
-\ln {\cal L'} = - \sum \ln p(v_z|R,r_\nu,\bmath{\eta}) \ .
\label{lnL2}
\end{equation}
Now, if $r_\nu$ is not known, then one
may be tempted to solve for it by minimizing $-\ln {\cal L}_0=-\sum \ln p_0(R|r_\nu)$, and
then proceed with equation~(\ref{lnL2}) to minimize for the $N\!-\!1$ remaining
parameters,  $\bmath{\eta}$. However, since $-\ln {\cal L} = -\ln {\cal
  L}' - \ln {\cal L}_0$ (from eqs.~[\ref{lnL}] and [\ref{sumsum}]), the most likely solution for
$\bmath{\theta}$ that minimizes $-\ln {\cal L}$ 
will not in general be that which minimizes at the same
time $-\ln {\cal L}'$ and $-\ln {\cal L}_0$.
Moreover, if one seeks to obtain the distributions of parameters
$\bmath{\eta}$ \emph{and} $r_\nu$ consistent with the MLE solution (for example with
Markov Chain Monte-Carlo techniques), the joint analysis of
equations~(\ref{qdef}) and (\ref{lnL}) is required.
On the other hand, if $r_\nu$ is known from other data, while the
current dataset is known to have a completeness, $C(R)$, that is a function of projected
radius, then one could indeed minimize $\-\ln {\cal L}'$ of
equation~(\ref{lnL2}). The proper solution is then to minimize $-\ln {\cal L}$
weighting the data points by the inverse completeness, i.e. minimizing 
\begin{equation}
-\ln {\cal L''} = - \sum_{i=1}^n {\ln q(R_i,v_{z,i}|\bmath{\theta}) \over C(R_i)}
\ .
\label{lnLwcomp}
\end{equation}

For computational efficiency, 
we perform the following tasks:
\begin{enumerate}
\item
For each run of parameters, we first compute
$\log \sigma_r(r_j)$ from equation~(\ref{sigmar2}) on a logarithmic grid of
$r_j$, and compute cubic-spline coefficients at these radii. 
Then, when we compute  
the LOS integral of equation~(\ref{g1}) for each $\left
(R_i,v_{z,i}\right)$, we 
evaluate
$\sigma_r(r)$
with cubic spline interpolation (in log-log space, using the cubic spline
coefficients determined at the start).

\item For simple anisotropy models, the exponential term in
  equation~(\ref{sigmar2}) is given by equations~(\ref{KoverK}) and
(\ref{exp2intbetaovert}). 

\item We terminate the LOS integration in equation~(\ref{qdef}) at roughly 15 virial
  radii,\footnote{The virial radii are loosely defined here as the radius
    where the mean density of the halo is 200 times the critical density of
    the Universe.}  $r_{\rm v}$,
  instead of infinity, as the Hubble flow pushes the velocities of the 
material beyond this distance to values over $3\,\sigma_v$ above the mean of
the system (see \citealp*{MBM10}, hereafter MBM10). The LOS integration varies only very
slightly with the number of virial radii, so as long as the virial radius is
correct to a factor of two, this choice of integration limit is not an issue.
\end{enumerate}

We now need to choose a model for the shape of the 3D velocity distribution.
While MAMPOSSt, can, in principle, be run with any model,
the simplest one is the (possibly anisotropic) Gaussian distribution, which we describe in
Sect.~\ref{gaussian} below. 

\subsection{Gaussian 3D velocity distributions}
\label{gaussian}
The simplest assumption for the 3D velocity distribution is that it is Gaussian:
\begin{equation}
f_v(v_r,v_\theta,v_\phi) = {1\over (2 \pi)^{3/2} \sigma_r
  \sigma_\theta^2} \exp \left [
-{v_r^2\over 2\sigma_r^2}
-{v_\theta^2+v_\phi^2\over 2\sigma_\theta^2}
\right ] \ ,
\label{fvgauss}
\end{equation}
where the velocity dispersions $\sigma_i$ are functions of $r$.
This Gaussian distribution assumes no streaming motions: e.g. no rotation,
and no mean radial streaming, which is adequate for $R_{\rm max}<r_{\rm v}$
in high-mass haloes (i.e. groups and clusters) and $R_{\rm max} < 4\,r_{\rm
  v}$ in galaxy-mass haloes
\citep{CPKM08}.
Inserting equation~(\ref{fvgauss}) into 
equation~(\ref{dNdvz}) and integrating over $v_\phi$ leads to
\begin{eqnarray}
h(v_z|R,r) \!\!\!&\!\!\!\!=\!\!\!\!& \!\!\!\!\int_{-\infty}^{+\infty} \!\!\!\! {1\over
  2\pi\sqrt{1\!-\!\beta}\,\sigma_r^2 }\,
\exp\left\{\!-{\left[(1\!-\!\beta)\,v_r^2+v_\theta^2\right]\over
  2\,(1\!-\!\beta)\,\sigma_r^2}\!\right\}  {\rm d}v_\perp .\nonumber\\
\label{hofvz2}
\end{eqnarray}
Calling $\theta$ the angle between the line-of-sight (direction $z$) and the
radial vector $\bmath r$, one has
\begin{eqnarray}
v_r &=& v_z\,\cos\theta + v_\perp\,\sin\theta \ , \\
v_\theta &=& - v_z\,\sin\theta + v_\perp\,\cos\theta \ ,
\end{eqnarray}
with which the integral over $v_\perp$ in equation~(\ref{hofvz2}) yields
a Gaussian distribution 
of LOS
velocities at point P:
\begin{equation}
h(v_z|R,r) 
= {1 \over \sqrt{2 \pi \sigma_z^2(R,r)}}\, \exp \left [-{v_z^2\over
    2\,\sigma_z^2(R,r)} \right ] \ ,
\label{hgauss}
\end{equation}
of squared dispersion 
\begin{equation}
\sigma_z^2(R,r) = \left [1-\beta(r) \left ({R\over r}\right)^2 \right ]\,\sigma_r^2(r) \ .
\label{sigmaz}
\end{equation}
The
integral of $h(v_z|R,r)$ along the LOS is obtained from
equations~(\ref{g1}) and (\ref{hgauss}):
\begin{eqnarray}
g(R,v_z) 
&\!\!\!\!=\!\!\!\!& 
\sqrt{2\over \pi}\,
\int_R^\infty  {r\,\nu\over \sqrt{r^2-R^2}}\,
{\left (1-\beta R^2/r^2 \right)^{-1/2}\over \sigma_r}
\nonumber \\
&\!\!\!\!\!\!\!\!& \qquad \quad \quad \times 
\exp \left
[-{v_z^2\over 2\,\left(1-\beta R^2/r^2 \right)\,\sigma_r^2}
\right ]\,\d r \ .
\label{ggaussian}
\end{eqnarray}
According to equations~(\ref{pofvzgivenR1}) and (\ref{ggaussian}), the probability of measuring a
velocity $v_z$ at given projected radius $R$ is
\begin{eqnarray}
p(v_z|R) &=& {g(R,v_z) \over \Sigma(R)} \nonumber \\
&=& 
{1\over \sqrt{2\pi}}\,
{
\int_0^\infty (\nu/ \sigma_z)\,\exp \left [-v_z^2/\left(2\,\sigma_z^2\right) \right]\,\d z
\over
\int_0^\infty \nu\,\d z 
} \ .
\label{pofvzgaussian}
\end{eqnarray}

We remind the reader that $\beta$ is a chosen function of $r$, $\nu$
is a function of $r$ given by equation~(\ref{rho}), while $\sigma_r$ is a
function of $r$ given by equation~(\ref{sigmar2}). 
For isotropic systems ($\beta=0$), equation~(\ref{ggaussian}) leads to a Gaussian
distribution of LOS velocities.
However, for anisotropic velocity tensors, the distribution of LOS
velocities will generally not be Gaussian (as \citealp{Merritt87} found when
starting from distribution functions instead of Gaussian 3D velocities).
Hence, the Gaussian nature of $h(v_z|R,r)$ is not equivalent to the popular
assumption that $g(R,v_z)$ 
is Gaussian on $v_z$: even if $h(v_z|R,r)$ is a Gaussian at point P, its
integral along the LOS is not Gaussian, unless $\beta=0$ and $\sigma_r$ is constant.

If one of the parameters to determine with  MAMPOSSt is the normalization of
the mass profile, one should \emph{not} be tempted in expressing the radii in terms
of the 
virial radius $r_{\rm v}$, the velocities in terms of the virial velocity
$v_{\rm v}$, 
the tracer densities in terms of what we wish (as they appear in both the
numerator and denominator of eq.~[\ref{pofvzgaussian}]). Doing so, 
equation~(\ref{pofvzgaussian}) becomes
\begin{eqnarray}
p(\widetilde v_z|\widetilde R) &\equiv& v_v\,p(v_z|R) \nonumber \\
&=&
{1\over \sqrt{2\pi}}\,
{
\int_{0}^\infty (\widetilde \nu/ \widetilde \sigma_z)\,\exp 
\left [-\widetilde v_z^2/\widetilde \sigma_z^2 \right]\,\d\widetilde z
\over
\int_{0}^\infty \widetilde \nu\,\d \widetilde z 
} \ ,
\label{pofvzgaussiannorm}
\end{eqnarray}
where the quantities with tildes are in virial units.
Equation~(\ref{pofvzgaussiannorm}) indicates that when one varies the
$r_{\rm v}$ (and the virial velocity in proportion as $v_{\rm v} =
\sqrt{2/\Delta}\,H_0\,r_{\rm v}$), 
the highest
probabilities are reached for the highest normalizations: $\widetilde v_z$
becomes very small, while $\widetilde \sigma_z$ is unaffected to first order.
This unphysical result is the consequence of using a parameter (the virial
radius) as part of the data variable. On the other hand, using
equation~(\ref{pofvzgaussian}), one sees that the highest probabilities
$p(v_z|R)$ are reached at intermediate values of the normalization.

Taking the second moment of the velocity distribution
of equation~(\ref{pofvzgaussian}) leads to the equation of anisotropic
projection yielding the LOS velocity dispersion, 
$\sigma_z(R)$:
\begin{eqnarray}
\Sigma(R)\,\sigma_z^2(R) 
&\!\!\!\!=\!\!\!\!& \int_{-\infty}^{+\infty} v_z^2 \,  g(R,v_z)\,\d v_z \nonumber \\
&\!\!\!\!=\!\!\!\!& \sqrt{2\over\pi}\,
\int_R^\infty {\nu\,r\,\d r\over \sigma_z(R,r)\,\sqrt{r^2-R^2}}
\nonumber \\
&\mbox{}& \qquad \times
\int_{-\infty}^{+\infty} v_z^2 \exp\left [-{v_z^2\over
     2\,\sigma_z^2(R,r)}\right ] {\rm d} v_z \nonumber \\
&\!\!\!\!=\!\!\!\!& 2\,\int_R^\infty \nu \sigma_r^2 
\left [1-\beta(r) {R^2\over r^2}\right]\,{r\,\d r\over \sqrt{r^2-R^2}}
\ .
\label{siglosgaussian}
\end{eqnarray}
Equation~(\ref{siglosgaussian}) recovers the equation of anisotropic
kinematic projection, first derived by \cite{BM82}.

If interlopers are removed with a velocity cut $v_{\rm cut}(R)$, then the
expression for $h(v_z|R,r)$ becomes
\begin{equation}
h(v_z|R,r) = \frac{\exp\left\{-v_z^2/\left[2
  \sigma_z^2(R,r)\right]\right\}}{\sqrt{2 \pi } 
    \sigma_z(R,r)\, {\rm erf}\left\{v_{\rm
      cut}(R)/\left[\sigma_z(R,r)\sqrt{2}\right]\right\}} \ .
\end{equation}

In summary, MAMPOSST with Gaussian 3D velocities computes likelihoods from
equations~(\ref{lnL}), (\ref{qdef}) or (\ref{qdef2}), 
(\ref{hgauss}), (\ref{sigmaz}), and (\ref{sigmar2}), in that order.


\section{Tests}
\label{tests}

\subsection{Simulated haloes}
\label{s:haloes}

\begin{table*}
\begin{center}
\tabcolsep=4pt
\caption{Properties of 11 cosmological haloes}
\begin{tabular}{rrccccccc}
\hline
\multicolumn{1}{c}{rank} & \multicolumn{1}{c}{ID} & \multicolumn{1}{c}{$r_{200}$} & $r_{\rm s}$ & $r_{\rm H}$ & $r_{\rm B}$ & ${\cal A}$ & $r_{\beta}$ &
${\cal A}_{\infty}$ \\
\hline
1 & 18667 & 0.789 & 0.179 & 0.401 & 0.117 & 1.14 & 0.276 & 1.33 \\
2 & 21926 & 0.842 & 0.123 & 0.342 & 0.085 & 1.34 & 0.050 & 1.73 \\
3 & 30579 & 0.890 & 0.189 & 0.443 & 0.120 & 1.33 & 0.053 & 1.69 \\
4 & 25174 & 0.956 & 0.144 & 0.377 & 0.099 & 1.23 & 0.162 & 1.36 \\
5 &  3106 & 1.010 & 0.297 & 0.661 & 0.166 & 1.05 & 2.384 & 1.09 \\
6 &  8366 & 1.076 & 0.434 & 0.819 & 0.249 & 1.11 & 0.689 & 1.29 \\
7 & 13647 & 1.151 & 0.227 & 0.536 & 0.151 & 1.19 & 0.265 & 1.41 \\
8 & 1131 & 1.174 & 0.197 & 0.499 & 0.133 & 1.18 & 0.352 & 1.34 \\
9 & 17283 & 1.298 & 0.505 & 1.009 & 0.277 & 1.04 & 0.727 & 1.05 \\
10 &  434 & 1.374 & 0.317 & 0.699 & 0.210 & 1.30 & 0.165 & 1.70 \\
11 & 5726 & 1.660 & 0.407 & 0.921 & 0.249 & 1.42 & 0.050 & 2.20 \\
\hline
 & Stack & 1.09$\pm$0.08 & 0.26$\pm$0.04 & 0.60$\pm$0.08 & 0.17$\pm$0.02 & 1.21$\pm$0.04 & 0.26$\pm$0.08  & 1.45$\pm$0.10 \\
\hline
\end{tabular}
\label{t:haloprops}
\end{center}

\parbox{\hsize}{Notes:
Properties obtained from fits to the particle data of 11 haloes. 
Cols.~1 and 2: cluster identification;
col.~3: virial radius $r_{200}$;
col.~4: scale radius ($=r_{-2}$) of the NFW mass density profile
(eq.~[\ref{e:nfw}]);
col.~5: scale radius ($=2\,r_{-2}$) of the Hernquist mass density profile
(eq.~[\ref{e:her}]);
col.~6: scale radius ($\simeq 0.657\,r_{-2}$) of the Burkert mass density
profile (eq.~[\ref{e:bur}]);
col.~7: mean anisotropy (${\cal A}=\sigma_r/\sigma_\theta$) within $r_{200}$;
col.~8: anisotropy radius with the ML anisotropy model;
col.~9: asymptotic anisotropy (${\cal A}_{\infty}=\sigma_r/\sigma_\theta$) at infinite  radius
with the T anisotropy model. Radii are in units of $h^{-1}\,\rm Mpc$. The
measured anisotropies do not incorporate streaming motions.
}
\end{table*}

To test MAMPOSSt, we use cluster-mass haloes 
extracted by
\citet{Borgani+04} from their large cosmological hydrodynamical
simulation performed using the parallel Tree+SPH {\small GADGET--2}
code of \citet{Springel+05}.  The simulation assumes a cosmological
model with $\Omega_0=0.3$, $\Omega_\Lambda=0.7$, $\Omega_{\rm
  b}=0.039$, $h=0.7$, and $\sigma_8=0.8$. The box size is $L=192\,
h^{-1}$ Mpc. The simulation used $480^3$ DM particles and
(initially) as many gas particles, for a DM particle mass of
$4.62 \times 10^9\, h^{-1} \rm M_\odot$.  The softening length was set to
$22.5 \, h^{-1}$ comoving kpc until $z=2$ and fixed afterwards
(i.e., $7.5\, h^{-1}$ kpc). The simulation code includes explicit
energy and entropy conservation, radiative cooling, a uniform
time-dependent UV background \citep{HM96}, the self-regulated hybrid
multi-phase model for star formation \citep{SH03}, and a
phenomenological model for galactic winds powered by Type-II
supernovae.

DM haloes were identified by \citet{Borgani+04} at
redshift $z=0$ with a standard Friends-of-friends (FoF)
analysis applied to the DM particle set, with linking length 0.15
times the mean inter-particle distance. After the FoF identification,
the centre of the halo was set to the position of its most bound
particle.  A spherical overdensity criterion was then applied to
determine, for each halo, our proxy for the virial radius, $r_{200}$, where the mean density
is 200 times the critical density of the Universe.

To
save computing time, we worked on a random subsample of roughly two
million particles among the $480^3$.  We have extracted 11 cluster-mass haloes
from these simulations, among which, ten are about
logarithmically spaced in virial radius, $r_{200}$, while the $11^{\rm th}$ halo
is the most massive in the entire simulation. 
Their properties are
listed in Table~\ref{t:haloprops}. We made no effort to omit
irregular haloes, but among the list of 12 irregular haloes out of 105
extracted by MBM10 from the same simulation, 2 are in our sample
(haloes 17283 and 434).
We list the characteristic radii
$r_{\rm s}, r_{\rm H}, r_{\rm B}$ of three models fitted by MLE to the mass density profiles
of the particle data (from 0.03 to
$1\,r_{200}$), namely:
\begin{enumerate}
\item the NFW density profile 
\begin{equation}
\rho(r) \propto r^{-1}\,\left(r+r_{\rm s}\right)^{-2}
\ ,
\label{e:nfw}
\end{equation} 
where $r_{\rm s} \equiv r_{-2}$ is the radius of slope $-2$ in the
mass density profile, related to the concentration
$c \equiv r_{200}/r_{-2}$;
\item the Hernquist density profile \citep{Hernquist90}
\begin{equation}
\rho(r) \propto r^{-1}\,\left (r+r_{\rm H}\right)^{-3}
\ ,
\label{e:her}
\end{equation} 
where $r_{\rm H}=2 \ r_{-2}$,
\item the Burkert density profile \citep{Burkert95}
\begin{eqnarray}
\rho(r) \propto \left (r+r_{\rm B}\right)^{-1}\,\left (r^2+r_{\rm B}^2\right)^{-1}
\ ,
\label{e:bur}
\end{eqnarray} 
where $r_{\rm B} \simeq 0.657 \ r_{-2}$.
\end{enumerate}
Denoting the scales $r_{\rm s}$, $r_{\rm H}$ and $r_{\rm B}$ by the generic
  $r_\rho$, 
the mass profiles of these models (required for eq.~[\ref{sigmar2}]) can be written
\begin{equation} 
M(r) = M(r_\rho)\,{\widehat M\left ({r/ r_\rho}\right) \over \widehat
  M(1)}\ ,
\end{equation}
where
\begin{equation}
\widehat M (x) \!=\! \left \{\!\!\!
\begin{array}{ll}
\displaystyle \ln(x+1)-{x\over x+1}\ , & \hbox{(NFW)} \\
\\
\displaystyle \left ({x\over x+1}\right)^2\ , & \hbox{(Hernquist)} \\
\\
\displaystyle \ln \left [(x\!+\!1)^2 (x^2\!+\!1)\right] - 2 \,\tan^{-1}\!x \,. & \hbox{(Burkert)} \\
\end{array}
\right .
\label{Mofrs}
\end{equation}

The NFW model has long been known to fit well the density profiles of
$\Lambda$CDM haloes \citep{NFW96}, and while \cite{Navarro+04} found that
Einasto models fit them even better, {MBM10} found that the NFW model
describes the outer LOS velocity dispersion profile of the DM component of their stacked cluster-mass
halo in \citeauthor{Borgani+04}'s hydrodynamical cosmological simulation
even (slightly)
better than the Einasto model. 
The Hernquist model differs from the NFW one because
it has a steeper logarithmic slope at large radii, $\gamma \equiv
{\rm d}\ln \rho / {\rm d}\ln r = -4$ rather than $-3$.  The
Burkert model, on the other hand, has the same asymptotic $\gamma=-3$
as the NFW model, but a core at the centre, $\gamma=0$, rather than a
cusp ($\gamma=-1$ in both the NFW and Hernquist models).

In Table~\ref{t:haloprops}, we also list the values of the parameters
characterizing different velocity-anisotropy models, namely:
\begin{enumerate}
\item the constant anisotropy model
  $\sigma_r/\sigma_\theta=(1-\beta)^{-1/2}={\cal A}$ (`Cst' model  
hereafter), where we assume spherical symmetry and therefore
$\sigma_\theta=\sigma_\phi$;
\item the model  (`ML' model hereafter) of \citet{ML05b};
\begin{equation}
\beta_{\rm ML}(r)={1\over 2}\, \, {r\over r+r_{\beta}} \ ,
\label{e:ML}
\end{equation}
characterized by the anisotropy radius $r_{\beta}$;
\item a generalization of the ML model, which is also
a simplified version of the model of \citet{Tiret+07}, isotropic
at $r=0$ and with anisotropy radius identical to $r_{-2}$ (hereafter called `T' model):
\begin{equation}
\beta_{\rm T}(r)=\beta_{\infty} \, {r \over r+r_{-2}} \ ,
\label{e:T}
\end{equation}
characterized by the anisotropy value at large radii,
$\beta_{\infty}$. In our T model, the anisotropy radius is set to 
the scale radius of the mass density profile. Note also that in the
following we provide the values of ${\cal A}_{\infty} \equiv
(\sigma_r/\sigma_\theta)_{\infty} = (1-\beta_{\infty})^{-1/2}$, rather
than $\beta_{\infty}$.
\end{enumerate}
Note that the ML and the T models used here are identical for
$\beta_{\infty}=0.5$ and $r_\beta=r_{-2}$. With these values, the ML
and T models provide a good fit to the average anisotropy profile of a
set of cluster-mass cosmological haloes (MBM10).  

\begin{figure}
\includegraphics[width=\hsize]{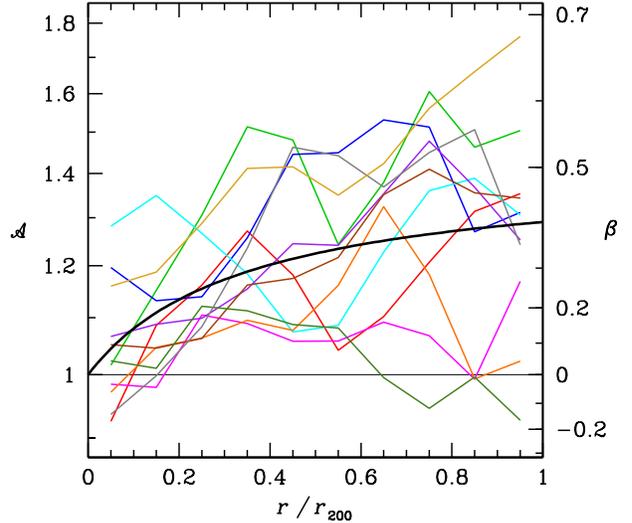}
\caption{Velocity anisotropy profiles of the 11 haloes (\emph{broken coloured
    lines}). 
The \emph{smooth black curve} is the ML anisotropy model with $r_{\beta}=r_{-2}$ (or,
  equivalently, the T anisotropy model with $\beta_{\infty}=0.5$).}
\label{f:betas}
\end{figure}

In
Fig.~\ref{f:betas}, we show the individual halo velocity anisotropy
profiles 
and
the ML anisotropy model with $r_{\beta}=r_{-2}$ (or, equivalently, the
T anisotropy model with $\beta_{\infty}=0.5$). There is a huge scatter
in the $\beta(r)$ of the individual haloes, as already observed by, e.g.,
\citet{Wojtak+08}, especially at $r > 0.3\,r_{200}$, while 8 out of 11
haloes have $\beta(0) = 0\pm0.15$.

\subsection{Observing cones and interloper removal}

To test MAMPOSSt, we select 500 DM particles around each
halo, out to a maximum projected distance  $R_{\rm
  max}$ from the halo centre, for which we consider three values:
$r_{500} \simeq 0.66\,r_{200}$, $r_{200}$, and $r_{100} \simeq 1.35\,r_{200}$.
We analyze three
orthogonal projections for each halo -- these are in fact cones with an observer at $D=90 \, h^{-1} \,
\mathrm{Mpc}$ away, but the opening angle being very small has no noticeable 
effect on our results.  The particles in these cones are used by MAMPOSSt as
tracers of the halo gravitational potential. 

However, these
500-particle samples include \emph{interlopers}, i.e. DM particles that
are located in projection at $R \leq R_{\rm max}$, but are effectively
outside $R_{\rm max}$ in real (3D) space, i.e. with $r>R_{\rm max}$.  
It is impossible to remove all
these interlopers in the observed redshift space, where only 3 of the
6 phase-space coordinates of the tracers are known (e.g., MBM10).  
Moreover, since
the LOS velocity distribution of interlopers in mock cones around
$\Lambda$CDM haloes is the sum of a Gaussian
component and a uniform one (see MBM10 for a quantified view), and since the Gaussian one
resembles that of the particles in the virial sphere, it is important
to remove the flat LOS velocity component, at least at high absolute
LOS velocity, where it dominates.  It is possible to remove these high
$|v_z|$ objects with suitable interloper removal algorithms.

To see how MAMPOSSt depends on the choice of the
interloper removal algorithm, we here consider three different
algorithms. 

The first one is a new, iterative algorithm, that we name
``Clean'', which is fully described in Appendix~\ref{clean}.
Clean first
looks for gaps in the LOS velocities,
then estimates the virial radius $r_{200}$ from the
aperture velocity dispersion, assuming an NFW model with ML anisotropy
with an anisotropy radius $r_{\beta} = r_{-2}$ and a concentration
depending on the estimate of $r_{200}$ via the relation of
\citet*{MDvdB08}, then only considers the galaxies within $2.7\,\rm
\sigma_z(R)$ from the median LOS velocity, and finally iterates.
Our assumed anisotropy profile 
fits
reasonably well the anisotropy profiles of DM haloes (MBM10), as is
clear for our 11 haloes (see Fig.~\ref{f:betas}).
The factor 2.7 was found by MBM10 to best preserve the local LOS
velocity dispersion for the assumed density and anisotropy models.

We also consider two other interloper removal algorithms, namely:
\begin{enumerate}
\item the method (hereafter, dHK) of \cite{dHK96}, a widely used
  procedure that works reasonably well on cluster-mass haloes from
  cosmological simulations \citep{Biviano+06,Wojtak+07}, despite its
  crude underlying physics;
\item the method (hereafter, KBM) of \citet*[][see their Appendix A]{KBM04},
in which a galaxy is flagged as an interloper under the condition 
$v_z /\sigma_z > 1.85\, (R/r_{200})^{-0.3}$, with $r_{200}$ derived from
$\sigma_z$ using eq.~(8) of \citet*{CYE97}. This method was invented
as a poor-man's proxy for the dHK method when the observational 
sampling of the halo projected phase-space is poor.
\end{enumerate}

\subsection{The general 4-parameter case}

\begin{figure*}
\centering
\includegraphics[width=0.8\hsize]{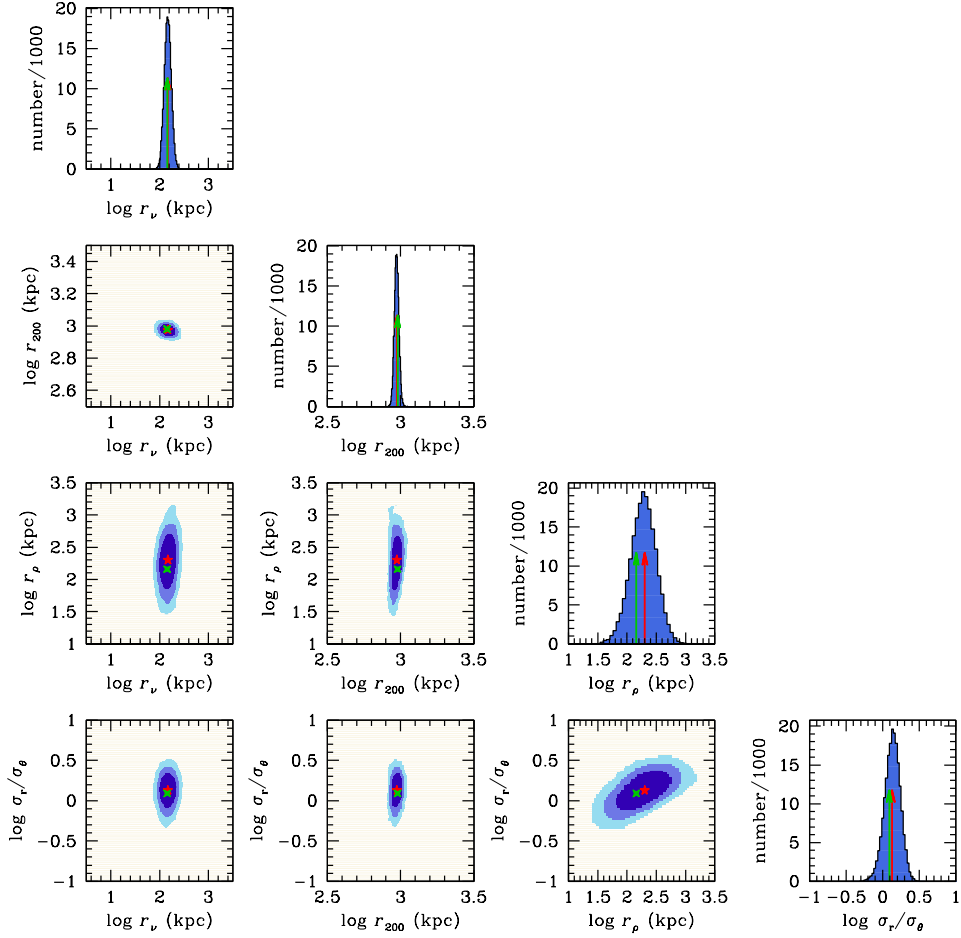} 
\caption{Illustration of MAMPOSSt analysis for the general case, with $\beta$
  independent of radius, for a 500 particle sample from axis $x$ of halo 25174
  (\emph{grey broken line} in Fig.~\ref{f:betas}), 
using MCMC (with 6 chains of $40\,000$ elements). 
The \emph{contours} are 1, 2, and $3\,\sigma$.
The \emph{red arrows} and \emph{stars} indicate the maximum likelihood
solution, while the \emph{green arrows} and \emph{crosses} show the true
solution (Table~\ref{t:haloprops}).
The priors for the MCMC were uniform within the boxes of each panel and zero
beyond the boxes.}
\label{figmcmc}
\end{figure*}

There is no {\em a priori} limitation on the number of free parameters
that can be used in MAMPOSSt to characterise the mass and velocity
anisotropy profiles. With samples of $\leq 500$ tracers (assumed massless
throughout these tests) 
it is
appropriate to consider $\sim 4$ free parameters, two for the mass
distribution, one for the velocity anisotropy distribution, and one
for the spatial distribution of the tracers. 
All these models are characterised by the two free
parameters, the `virial' radius $r_{200}$ and a characteristic scale-radius 
($r_{\rm s}$, $r_{\rm H}$,
and $r_{\rm B}$ for the NFW, Hernquist, and Burkert models,
respectively). Herafter, we generically use $r_{\rho}$ to refer to this
characteristic scale-radius of the mass density profile.

We use the NFW model, in projection \citep{Bartelmann96,LM01}, to fit
the projected number density profile of the tracer. Note that the
normalization of this profile does not enter the MAMPOSSt equations,
so the only free parameter is $r_{-2}$. Herafter we call
this parameter  $r_{\nu}$, 
to avoid confusion with the characteristic radius of
the NFW mass density profile. We only consider one model for the
number density profile of the tracer, because this is a direct
observable, unlike the mass density profile. While 
one should not
be too restrictive in the model choice for the mass density
profile, the observer is generally able to choose the best-suited
model for the tracer number density profile by direct examination of
the data before running MAMPOSSt. We choose the NFW model because it
provides a reasonable description of the number density profiles of
the DM particles in our simulated haloes.

For the velocity anisotropy profile, we consider the three models
described above, Cst, ML, and T, each characterised by a single
anisotropy parameter, ${\cal A}, r_{\beta},$ and  ${\cal A}_{\infty}$,
respectively. In equation~(\ref{e:T}), we use $r_{-2}=r_\rho$.

To search for the best-fit solution, we run the MAMPOSSt algorithm in
combination with the NEWUOA\footnote{NEWUOA is available at
  http://plato.asu.edu/ftp/other\_software/newuoa.zip} 
minimization routine of \cite{Powell06}.
For estimating error bars on the best fit parameters, as well as confidence
contours on pairs of parameters, 
we fit our model parameters using the Markov Chain Monte Carlo (MCMC)
technique (e.g., \citealp{LB02}).
In MCMC, the $k$-dimensional parameter space is populated with
\emph{proposals}, for each of which the likelihood is computed.
The new proposal is accepted if the 
ratio of new to previous likelihood is either greater than unity or else
greater than a uniform $[0,1]$ random number.
The proposal is found by 
assuming a $k$-dimensional Gaussian probability distribution around the
previous proposal.
We adopt the publicly available CosmoMC code by A. Lewis.\footnote{CosmoMC is
  available at {\tt http://cosmologist.info/cosmomc/}.}
We run 6 chains in parallel using Message Parsing Interface (MPI), and the
covariance matrix is used to update the parameters of the  Gaussian proposal
density to ensure faster convergence.

Fig.~\ref{figmcmc} illustrates the MAMPOSSt analysis via MCMC for
the general case with four free parameters, using the NFW model for the
mass density profile, and the constant (free parameter) anisotropy model.
In particular, it shows that the different parameters are
not correlated, except for a positive correlation between $r_\rho$ and
${\cal A}$.

Our FORTRAN code takes roughly 1~second to find the
MAMPOSSt 4-parameter solution for a 500~particle sample run in scalar on a
decent desktop or laptop computer, and 4~minutes to
produce confidence limits for this solution with the CosmoMC \citep{LB02}
MCMC code, with 6 chains of $40\,000$ elements run in parallel (MPI) on a
PC equipped with a 4-core 8-thread Intel Core-I7 2600 processor.

\begin{table*}
\caption{MAMPOSSt results for different interloper removal algorithms,
  density models, apertures, and number
  of particles}
\label{t:gen}
\tabcolsep=4pt
\begin{tabular}{lccccrrrcrrrcrrrcrrr}
\hline
\hline
N & $R_{\rm max}$ & Membership & $\rho(r)$ & $\beta(r)$ & \multicolumn{3}{c}{$r_{200}$} & &
\multicolumn{3}{c}{$r_{\nu}$} &  & \multicolumn{3}{c}{$r_{\rho}$} & & \multicolumn{3}{c}{anisotropy} \\
\cline{6-8}
\cline{10-12}
\cline{14-16}
\cline{18-20}
& & & & & bias & ineff. & corr. & & bias & ineff. & corr. & & bias & ineff. &
corr. & & bias & ineff. & corr. \\
\hline
500 & $r_{200}$ & Clean &  NFW & Cst &  0.004 & 0.040 & {\bf  0.909} & &  0.027 & 0.102 & {\bf  0.835} & &  0.032 & 0.217 & {\bf  0.578} & &   0.007 & 0.073 &          --0.255  \\
500 & $r_{200}$ & Clean & NFW &  ML & --0.003 & 0.040 & {\bf  0.904} & &  0.024 & 0.104 & {\bf  0.832} & &  0.057 & 0.229 & {\bf  0.601} & & --0.221 & 0.887 &          --0.172  \\
500 & $r_{200}$ & Clean & NFW &   T & --0.006 & 0.040 & {\bf  0.903} & &  0.026 & 0.103 & {\bf  0.838} & &  0.039 & 0.169 & {\bf  0.709} & & 0.007 & 0.085 & {\bf  0.621} \\
\hline
500 & $r_{200}$ &   dHK & NFW & Cst &  0.018 & 0.042 & {\bf  0.885} & &  0.027 & 0.099 & {\bf  0.838} & &  0.051 & 0.319 &           0.406 &  &  0.004 & 0.147 &          --0.215  \\
500 & $r_{200}$ &   dHK & NFW &  ML &  0.012 & 0.041 & {\bf  0.909} & &  0.028 & 0.100 & {\bf  0.840} & &  0.059 & 0.286 & {\bf  0.611} & & --0.161 & 0.904 &          --0.118  \\
500 & $r_{200}$ &   dHK & NFW &   T &  0.012 & 0.044 & {\bf  0.902} & &  0.027 & 0.100 & {\bf  0.844} & &  0.022 & 0.199 & {\bf  0.636} & & --0.045 & 0.264 & {\bf  0.464} \\
\hline
500 & $r_{200}$ &   KBM & NFW & Cst &  0.005 & 0.038 & {\bf  0.906} & &  0.018 & 0.100 & {\bf  0.850} & & --0.006 & 0.218 & {\bf  0.535} & &  0.020 & 0.078 &          --0.198  \\
500 & $r_{200}$ &   KBM & NFW &  ML & --0.003 & 0.039 & {\bf  0.908} & &  0.020 & 0.100 & {\bf  0.851} & & --0.005 & 0.232 & {\bf  0.557} & & --0.191 & 0.795 &           0.101  \\
500 & $r_{200}$ &   KBM & NFW &   T & --0.006 & 0.038 & {\bf  0.911} & &  0.020 & 0.099 & {\bf  0.856} & & --0.010 & 0.184 & {\bf  0.689} & & 0.018 & 0.094 & {\bf  0.595} \\
\hline
500 & $r_{200}$ & Clean & Her &   T &   0.002 & 0.039 & {\bf  0.909} & &  0.026 & 0.102 & {\bf  0.835} & &  0.039 & 0.132 & {\bf  0.755} & & 0.014 & 0.086 & {\bf  0.546} \\
500 & $r_{200}$ & Clean & Bur &   T &   0.000 & 0.039 & {\bf  0.910} & &  0.047 & 0.196 & {\bf  0.377} & &  0.048 & 0.145 & {\bf  0.704} & & --0.019 & 0.071 & {\bf  0.603} \\
\hline
500 & $r_{500}$ & Clean &  NFW & T & --0.004   & 0.048 & {\bf  0.877} & & 0.089  & 0.115 & {\bf 0.902} & & 0.016   & 0.143 & {\bf 0.744}      & & 0.009  & 0.108 & 0.232             \\
500 & $r_{100}$ & Clean &  NFW & T & --0.014 & 0.035 & {\bf  0.905} & & 0.039  & 0.179 & {\bf 0.420} & & 0.093   & 0.210 & {\bf 0.538}      & & --0.001  & 0.090 & 0.436       \\
\hline
100 & $r_{200}$ & Clean &  NFW & Cst & --0.001 & 0.058 & {\bf  0.844} & &  0.033 & 0.201 & {\bf  0.537} & & --0.133 & 0.341 &           0.342  & &  0.003 & 0.119 &          --0.053  \\
100 & $r_{200}$ & Clean & NFW &  ML & --0.011 & 0.061 & {\bf  0.834} & &  0.034 & 0.199 & {\bf  0.539} & & --0.087 & 0.336 & {\bf  0.484} & & --0.137 & 0.925 &          --0.245  \\
100 & $r_{200}$ & Clean & NFW &    T & --0.008 & 0.058 & {\bf  0.850} & &  0.032 & 0.200 & {\bf  0.532} & & --0.108 & 0.277 &           0.436 &  &  0.014 & 0.143 &           0.249  \\
\hline
\hline
\end{tabular}

\parbox{\hsize}{Notes: 
These results are for 11 haloes each observed along 3 axes,
  general 4 free-parameter case.
Col.~1: Number of initially selected particles (before interloper removal);
col.~2: Maximum projected radius for the selection, where $r_{500} \simeq
0.65\,r_{200}$ and $r_{100} \simeq 1.35 r_{200}$;
col.~3: Interloper-removal method 
(dHK: \citealp{dHK96}; 
 KBM: \citealp{KBM04}; 
 Clean: App.~\ref{clean}); 
col.~4: mass density model (NFW: \citealp{NFW96}; Her: \citealp{Hernquist90};
Bur: \citealp{Burkert95}); 
col.~5: anisotropy model (Cst: $\beta=\rm cst$; ML:
  eq.~[\ref{e:ML}], \citealp{ML05b}; T: eq.~[\ref{e:T}], adapted from \citealp{Tiret+07});
cols.~6--8: virial radius;
cols.~9--11: tracer scale radius;
cols.~12--14: dark matter scale radius;
cols.~15--17: velocity anisotropy (i.e., ${\cal A}$ for the Cst model, $r_{\beta}$
for the ML model, and ${\cal A}_{\infty}$ for the T model).
The columns `bias' and `ineff.' respectively provide the mean and standard
deviation (both computed with the biweight  technique) of $\log (o/t)$, while
columns `corr.' list the 
Spearman rank correlation coefficients between the true values and  MAMPOSSt-recovered ones
(values
in boldface indicate significant correlations between $o$ and $t$
values at the $\geq 0.99$ confidence level).}
\end{table*}

\begin{figure}
\centering
\includegraphics[width=\hsize]{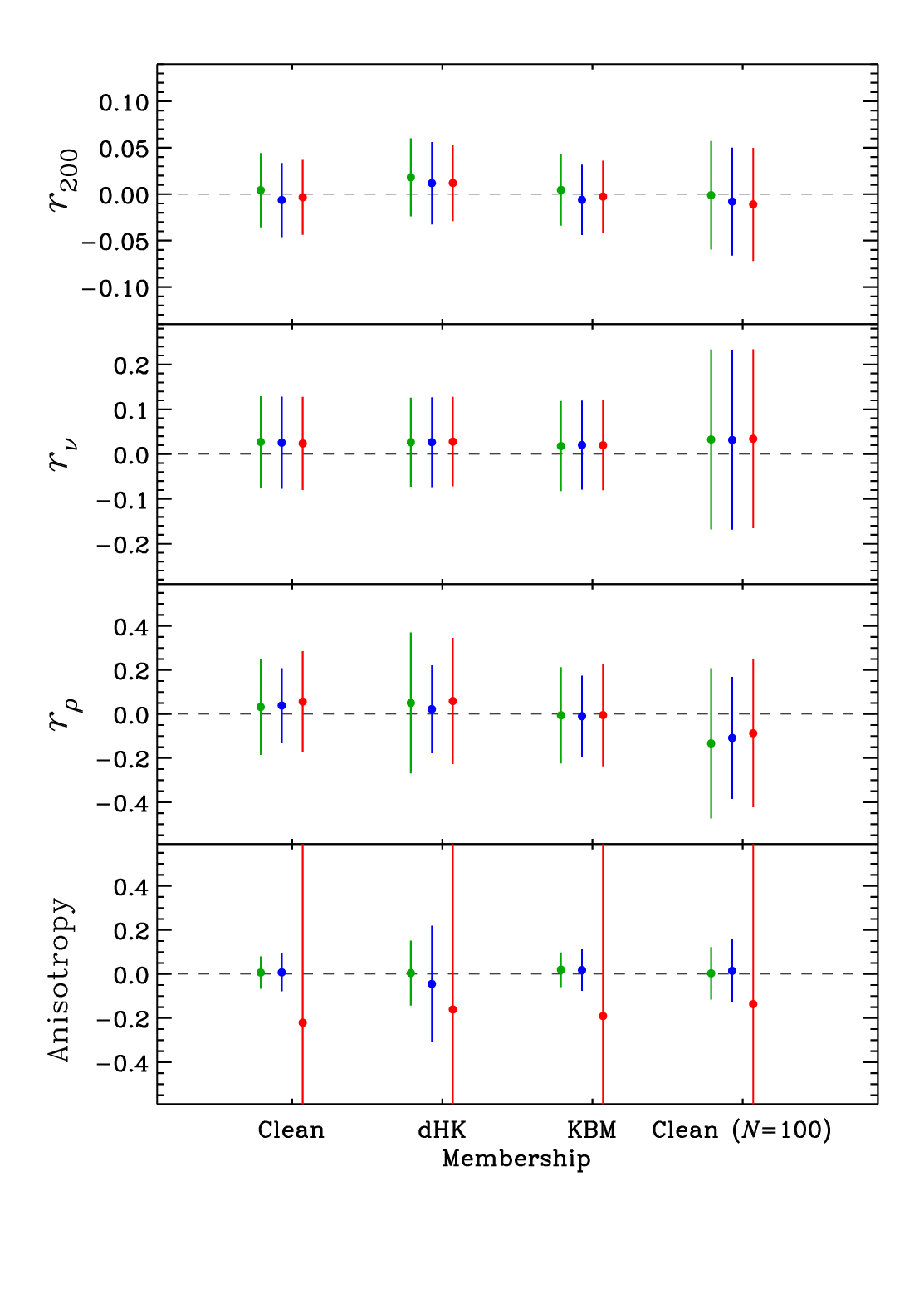}
\caption{MAMPOSSt residuals, $\log (o/t)$, for the MAMPOSSt
parameters (\emph{top}: virial radius, \emph{2nd panel}: tracer scale radius,
\emph{3rd panel}: DM scale radius, \emph{bottom}: velocity anisotropy)
for the different schemes of interloper removal (see text). 
The mean (\emph{dots}) and dispersion (\emph{error bars}) of $\log(o/t)$
are respectively illustrated  as \emph{filled circles} and \emph{error bars},
   for the 33 samples of  500 (`Clean', dHK and KBM) and 100 (`Clean' ($N$=100))
  particles.
  Results for the anisotropy models Cst, T, and ML are
  shown left to right in \emph{green}, \emph{blue,} and \emph{red},
  respectively.}
\label{f:gen}
\end{figure}

The results for the different interloper rejection methods, mass
density and velocity anisotropy models, and for the different maximum
projected 
radii used in the selection of the 500 particles are listed in
Table~\ref{t:gen} and displayed in Fig.~\ref{f:gen}. 
We list and show
the biweight measures \citep*[see, e.g.,][]{BFG90} of mean and
dispersion of $\log (o/t)$ where $o$ is the recovered value of the
parameter and $t$ its true value, because, according to our tests, 
they perform better than standard
statistical estimators of location and scale when the parent distributions
are not pure Gaussians. We call `bias' and `inefficiency'
the mean and dispersion of $\log (o/t)$. If the dispersion in true
values of a given parameter is small, one can spuriously obtain low
values of the $\log(o/t)$ dispersion when the MAMPOSSt and true values
show no correlation.  We therefore also list the Spearman rank
correlation coefficient between $o$ and $t$, marking in boldface those
correlations that are significant at the 99\% confidence level. We
list the results for all interloper rejection methods and all
anisotropy models only for the NFW mass density model and for the $R
\leq r_{200}$ radial selection. For simplicity, we only show a limited
set of results for the other mass density models and for the other
radial selections.

Remarkably, as seen in Table~\ref{t:gen}, the results for the four
parameters are almost independent of the interloper removal algorithm,
the Clean and KBM algorithm performing slightly better than
dHK. The
results for $r_{200}, r_{\nu},$ and $r_{\rho}$ also depend very little
on the chosen anisotropy model.  On average, the values of the
$r_{200}$ parameter are recovered with almost no bias (from $-1$ to
$+4$\%) and with only $\sim 10$\% inefficiency. The $r_{\nu}$
parameter estimates are always slightly positively biased (4--7\%),
and are recovered with $\sim 25$\% efficiency. Also, the $r_{\rho}$
parameter estimates generally display a slight positive bias, except
for the KBM interloper removal method, and overall the bias ranges
from $-2$ to $+15$\%, while the efficiency ranges from $\sim 50$ to
$\sim 90$\%.

As far as the anisotropy parameter is concerned, the ML model behaves
very differently from the Cst and T models, in that it is virtually
impossible to constrain the anisotropy radius of the former,
$r_{\beta}$, while it is possible to obtain quite good constraints on
the anisotropy parameters of the other two models, ${\cal A}$ and
${\cal A}_{\infty}$. More precisely, the $r_{\beta}$ estimates are
always negatively biased (by $\sim 60$\%) and are affected by a huge
dispersion (almost one order of magnitude). On the other hand, ${\cal
  A}$ and ${\cal A}_{\infty}$ are almost unbiased (the bias ranges
from $-10$ to $+5$\%) and they are affected by dispersions of,
typically, $\sim 20$\%, if we consider the Clean and KBM interloper
removal algorithms.  So, apparently, it is much easier to constrain
the `normalisation' of a given anisotropy profile, than to constrain
the characteristic radius at which the anisotropy changes,
particularly so if this change is mild, as in the ML model. Note,
however, that the difficulty of MAMPOSSt in constraining the anisotropy
parameter of the ML model does not mean that the ML model is a poor
representation of reality, and in fact Fig.~\ref{f:betas} suggests the
opposite. Moreover, constraints obtained on the $r_{200}, r_{\nu},$
and $r_{\rho}$ parameters are equally good with the Cst and T
anisotropy models, as with the ML one.

As seen in Table~\ref{t:gen}, 
correlations between recovered and observed values of the parameters
$r_{200}, r_{\nu},$ and $r_{\rho}$ are almost always signficant. This
is also true for the ${\cal A}_{\infty}$ anisotropy parameter, but not
for ${\cal A}$ and $r_{\beta}$.  
\begin{figure}
\includegraphics[width=\hsize]{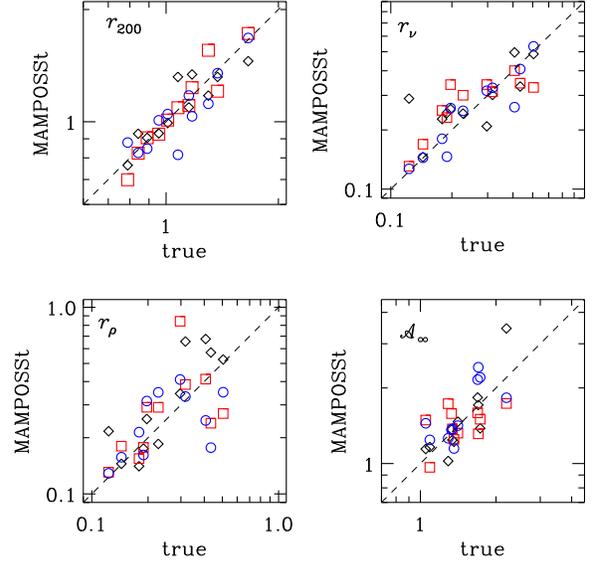}
\caption{Correlation of MAMPOSSt and true values of the 4 jointly-fit
  parameters (Case Gen), with the `T' anisotropy profile, for each of the
  $3\times11$ haloes with 500
  tracers. Each panel corresponds to a different parameter, as
  labelled (units of radii are in $h^{-1}$ Mpc). Different symbols identify
  different projections, $x$-axis: 
  \emph{black diamonds}, $y$-axis: \emph{red squares}, $z$-axis: \emph{blue circles}.}
\label{f:corr}
\end{figure}
In Fig.~\ref{f:corr}, we show the
correlations existing between the true and recovered values of the
different parameters, using the T anisotropy model, for the 11 haloes
along the 3 different orthogonal projections. Projections effects render
the determination of the mass and anisotropy profile of a single
500-particle halo very uncertain.  However, Fig.~\ref{f:corr} shows that $\sim 500$
tracers are sufficient to rank haloes for each of the different
parameters considered here, i.e. by mass ($r_{200}$), scale radius  of
the tracer distribution ($r_{\nu}$) and of the total mass distribution
($r_{\rho}$), and outer velocity anisotropy ${\cal A}_{\infty}$.

\begin{figure}
\includegraphics[width=\hsize]{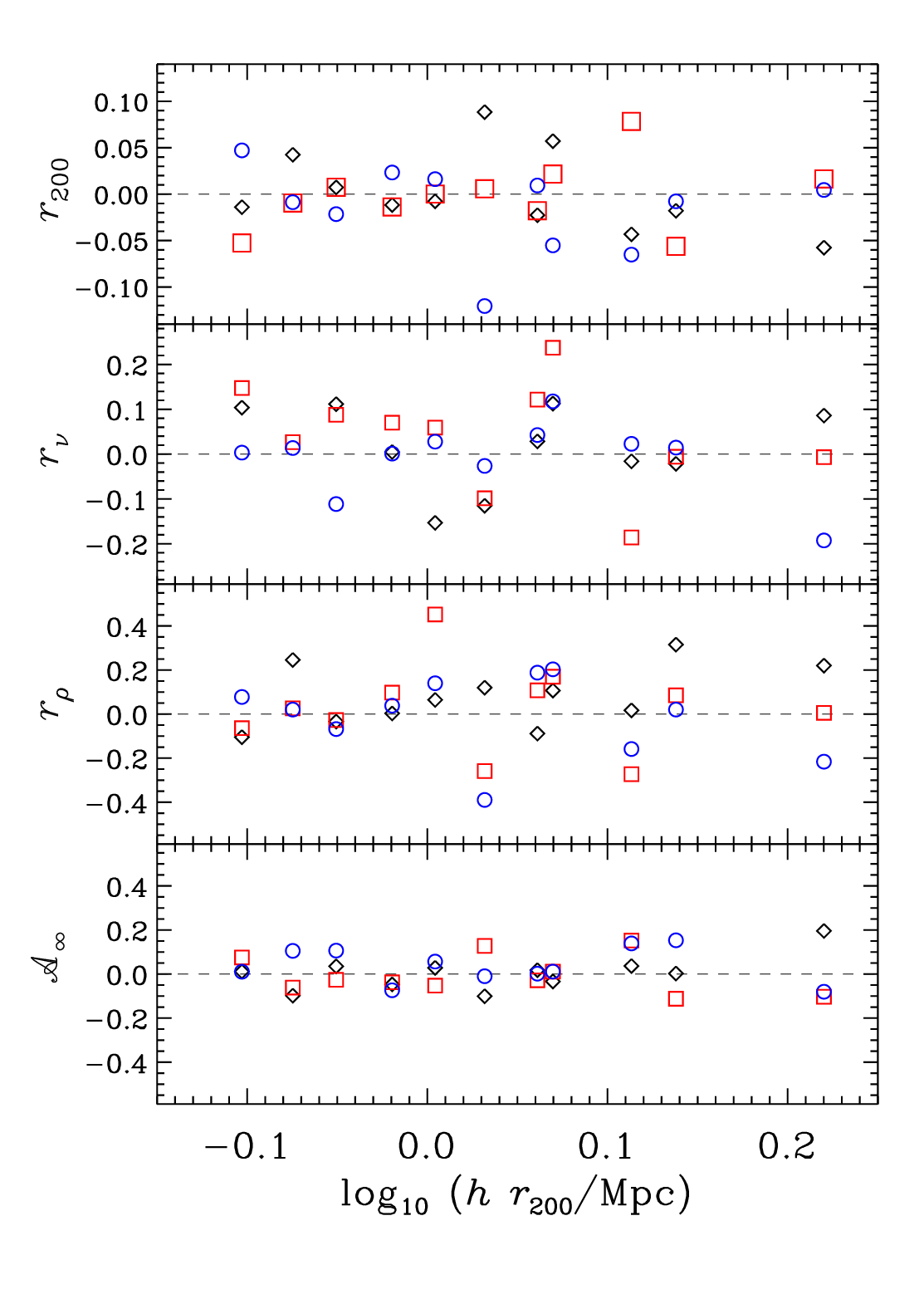}
\caption{MAMPOSSt residuals as a function of true virial radius for the 
4 jointly-fit
  parameters (Case Gen), with the `T' anisotropy profile, for each of the
  $3\times11$ haloes with 500
  tracers. Each panel corresponds to a different parameter, as
  labelled.
  Different symbols identify different
  projections,  $x$-axis:
  \emph{black diamonds}, $y$-axis: \emph{red squares}, $z$-axis: \emph{blue
    circles}.}
\label{f:anist}
\end{figure}
The importance of projection effects is also very clear from
Fig.~\ref{f:anist}, where we display the ratio of the recovered to
true values of the parameters for each halo along the three different
projection axes. This figure also shows that there is no trend
of under- or over-predicting the parameter values with halo mass.

All the above considerations apply for the NFW mass profile. Our tests
with the Hernquist and Burkert mass profiles give similar results, as
can be seen in Table~\ref{t:gen}, where for simplicity, we only list
the results for the Clean interloper-removal algorithm and for the
Tiret anisotropy model. The results are very similar for the NFW and
Hernquist models. Results are similar also for the Burkert model,
except for the scale $r_\nu$ of the number density profile, for which the bias and
inefficiency are both higher than those
obtained using the NFW and Hernquist mass models. 
Since the model we use for the number density profile has not
changed (a projected NFW), this result suggests that it is
difficult to accomodate a tracer with a central cuspy spatial
distribution in a potential with a central core.

All the results described so far were obtained for a selection of 500
particles within $R_{\rm max} = r_{200}$. Changing the value of
$R_{\rm max}$ is not without effects on the results. The inefficiency on
$r_{200}$ decreases when $R_{\rm max}$ is gradually 
increased from $r_{500}$ to $r_{100}$. The inefficiency on
anisotropy is largest for the smallest $R_{\rm max}$, and statistically
similar for the two larger values.
Increasing the aperture to the virial radius or above
increases the number of tracers
near the virial radius where $r_{200}$ is estimated. Moreover, increasingly
larger apertures will capture better the 
asymptotic value of the anisotropy profile ${\cal A}(r)$.
In contrast, 
$r_{\rho}$ is less efficiently determined when 
$R_{\rm max}$ is increased from $r_{500}$
to $r_{100}$, while $r_\nu$ has its worst inefficiency for $R_{\rm max}=r_{100}$,
with statistically equivalent values for the two smaller apertures. 
This might be due to the increasing
fraction of unidentified interlopers, and/or to the presence of
(sub)structures at larger radii.  

To assess the sensitivity of the MAMPOSSt technique to the number of
tracers, besides the 500-particle selection, we have also considered
samples of 100 particle tracers, randomly extracted from the same
projections of the same 11 cosmological haloes. Results of the MAMPOSSt
analysis are listed at the bottom of Table~\ref{t:gen} and displayed
in Fig.~\ref{f:gen}. Also in this case, for simplicity, we only
display a limited set of results.  When compared to the results for
the 500-particle samples, there is no significant change in the
average values of the bias with which the different parameters are
recovered, except for $r_{\rho}$, where the bias becomes negative,
while it was mostly positive for the 500-particle samples. The
$r_{\rho}$ parameter value underestimation is not very severe,
however, $\leq 25$\%. The efficiencies with which the different
parameters are estimated are significantly affected by the reduction
in number of tracers. The dispersion increases from $\sim 10$ to 15\%
for $r_{200}$, from $\sim 25$ to 60\% for $r_{\nu}$, from $\sim 60$
to 100\% for $r_{\rho}$, and from $\sim 20$ to 33\% for ${\cal A}$
and ${\cal A}_{\infty}$. There is no significant change in the
dispersion for $r_{\beta}$, but this was already extremely large for
the 500-particle samples.

\subsection{Cases with constraints on parameters}
\label{s:testothers}
What is the effect of reducing the number of free parameters on the
performance of the MAMPOSSt algorithm? To assess this point we
consider several cases that reproduce what observers do in practice
when faced with the problem of determining the internal dynamics of
cosmological haloes.  In all cases, we consider 500 particles selected
within $r_{200}$ in each halo. We only apply the Clean interloper
removal algorithm, and we only consider the NFW mass density model,
for simplicity.

Specifically, we consider the following Cases:
\renewcommand{\theenumi}{\Alph{enumi})}
\begin{enumerate}
\item General [Gen]: $r_{200}$, $r_{\nu}$, $r_{\rho}$, and the
  anisotropy parameter (one among the following: ${\cal A}$,
  $r_{\beta}$, ${\cal A}_{\infty}$, depending on the anisotropy model
  considered) are all free MAMPOSSt parameters. This is the case
  considered so far.
\item General with $r_{\nu}$ fitted outside MAMPOSSt [Split]: the free
  parameters are the same as in the Gen case, but $r_{\nu}$ is fitted
  outside MAMPOSSt, via MLE. We thus split
  the minimization of the parameters into two parts.
\item Known virial mass or radius [KVir]: $r_{200}$ is fixed and assumed to be exactly
  known, $r_{\rho}$ and the anisotropy parameter are free parameters
  in MAMPOSSt, $r_{\nu}$ is an external free parameter, as in the
  Split case.
\item Estimated virial mass or radius [EVir]: similar to KVir, except that $r_{200}$
 is not the true value, but the value estimated from the LOS
 aperture velocity dispersion (after interloper removal, see Appendix~\ref{clean}).
\item $\Lambda$CDM: $r_{\rho}$ is estimated from $r_{200}$ using
the theoretical relation between these two quantities provided
by \citet{MDvdB08}; $r_{200}$ and the anisotropy parameter are free parameters
in MAMPOSSt, $r_{\nu}$ is an external free parameter, as in the
Split case.
\item Mass follows Light [MfL]: $r_{200}$ and the anisotropy parameter
  are free parameters in MAMPOSSt, $r_{\nu}$ is an external free
  parameter, as in the Split case, and $r_{\rho}$ is assumed to be
  identical to $r_{\nu}$.
\item Tied Light and Mass [TLM]: $r_{200}$ and the anisotropy
  parameter are free parameters in MAMPOSSt, while $r_{\rho}$ and  $r_{\nu}$
  are assumed to
  be an identical free parameter.
\item Isotropic [$\beta$-iso]: ${\cal A}=1$ is assumed, $r_{200}, r_{\rho}$ are free
  parameters in MAMPOSSt, $r_{\nu}$ is an external free parameter, as
  in the Split case.
\item Anisotropy model \`a la {MBM10} [$\beta$-MBM]: ML anisotropy model
with $r_{\beta}$ forced to be identical to $r_{\rho}$;
  $r_{200}, r_{\rho}$ are free parameters in MAMPOSSt, $r_{\nu}$ is an
  external free parameter, as in the Split case.
\item Anisotropy linked to the mass density profile, using the anisotropy -
  slope relation
of \citet{HM06} [$\beta$-HM].
\end{enumerate}

\begin{table}
\begin{center}
\caption{Parameter inputs for the different cases of parameter constraints}
\label{t:cases}
\begin{tabular}{lcccc}
\hline
\hline
Case & $\Sigma(R)$ & \multicolumn{2}{c}{$M(r)$} & $\beta(r)$ \\
\cline{3-4}
& $r_\nu$ & $r_{200}$ & $r_\rho$ & \\
\hline
Gen & $\{R,v_z\}$ &  $\{R,v_z\}$ &  $\{R,v_z\}$ &  $\{R,v_z\}$ \\
Split & $\{R\}$ &  $\{R,v_z\}$ &  $\{R,v_z\}$ &  $\{R,v_z\}$ \\
KVir & $\{R\}$ & known &  $\{R,v_z\}$ &  $\{R,v_z\}$ \\
EVir & $\{R\}$ & from $\sigma_z$ &  $\{R,v_z\}$ &  $\{R,v_z\}$ \\
$\Lambda$CDM &  $\{R\}$ &  $\{R,v_z\}$ &  from $r_{200}$  &  $\{R,v_z\}$ \\
MfL & $\{R\}$ &  $\{R,v_z\}$ &  $r_\nu$ &  $\{R,v_z\}$ \\
TLM & $r_\rho$ &  $\{R,v_z\}$ &  $\{R,v_z\}$ &  $\{R,v_z\}$ \\
$\beta$-iso & $\{R\}$  &  $\{R,v_z\}$ &  $\{R,v_z\}$ & 0 \\
$\beta$-MBM & $\{R\}$  &  $\{R,v_z\}$ &  $\{R,v_z\}$ & $r_\beta$=$r_\rho$ \\
$\beta$-HM & $\{R\}$  &  $\{R,v_z\}$ &  $\{R,v_z\}$ & $a + b\,\d\ln\nu/\d\ln r$ \\
\hline
\end{tabular}
\end{center}
\end{table}
Table~\ref{t:cases} summarizes the different Cases, indicating, for
each parameter, whether it is estimated from the full $\{R,v_z\}$
distribution with MAMPOSSt, the $\{R\}$ distribution only,
using standard MLE, or if it is fixed or linked to
some other parameter.

The Cases outlined above correspond to different observational
situations.  Case Gen is the most general situation in which the
observer ignores all the dynamical characteristics of the system, and
it is the case analysed so far (see Table~\ref{t:gen}).  We repeat it
here for the sake of comparison with the other cases.  

Case Split is
closer than case Gen to the typical observational situation, since the
number density profile of the tracers of the potential (stars,
galaxies) is generally determined directly (e.g. by counting them in
concentric radial annuli, if the system is assumed to be spherical).
The advantage of Split relative to Gen is that any radial
incompleteness that might affect the determination of $r_{\nu}$ can be
easily corrected for, if known, before running MAMPOSSt \citep[see,
  e.g.][]{BP09}. However, as outlined in Sect~\ref{genmeth}, an interesting 
alternative is to include this incompleteness as in equation~(\ref{lnLwcomp}).
Here,
in all other cases (except TLM),
$r_{\nu}$ has been determined outside MAMPOSSt. Anyway, the results
(see below) obtained for the Gen and Split cases are very similar,
hence it makes little difference if $r_{\nu}$ is fitted within or
outside MAMPOSSt, when the analysed sample (as in our case) does not
suffer from radial-dependent incompleteness.

In the KVir and EVir cases, $r_{200}$ is not a free parameter, but 
is fixed externally. In the KVir case, a perfect knowledge of $r_{200}$
is assumed, and in fact we take the true $r_{200}$ of the cosmological
haloes. This case corresponds to the situation in which 
$r_{200}$ estimates are available from other data than the
projected-phase space distribution of galaxies, e.g. from weak-lensing
or X-ray observations for clusters, 
although in the real world also these mass
estimates will be affected by uncertainties, so KVir is a rather
idealised case.  Case EVir corresponds to the situation in which
$r_{200}$ estimates are directly obtained from the Clean interloper
removal scheme, as
described in Appendix~\ref{clean}.

In the $\Lambda$CDM and MfL cases, $r_{\rho}$ is not a free parameter. In the
$\Lambda$CDM case, we determine $r_{\rho}$ from $r_{200}$ using the relation of
\citet{MDvdB08}, which is based on the analysis of haloes in
$\Lambda$CDM numerical simulations.  Case MfL corresponds to the
situation in which the observer has good reasons (or {\em a priori}
theoretical prejudice) to assume that `Mass follows Light', i.e. that
the tracer is spatially distributed like the mass. Therefore the
observer first determines $r_{\nu}$ from the distribution of the
tracer and then makes the assumption $r_{\rho} = r_{\nu}$. In case
TLM, the observer is unable to
constrain the tracer scale radius from its spatial distribution. This may happen when dealing with an incomplete sample with
unknown incompleteness. Hence, $r_{\nu}$ and $r_{\rho}$ are
both determined from the dynamical analysis, assuming they are identical.

In the $\beta$-iso, $\beta$-MBM, and $\beta$-HM cases, the anisotropy
is no longer a free parameter.  The $\beta$-iso case corresponds to
the situation where, for lack of better knowledge, the velocity
anisotropy profile is assumed to be isotropic, ${\cal A} = 1$
($\beta=0$) \citep[like in, e.g.][]{BG03}. Since the velocity
anisotropy profiles of cluster-mass haloes have been shown to be well
represented by an ML profile with $r_{\beta} = r_{\rho}$
(MBM10, see also Fig.~\ref{f:betas}), it also makes sense to
fix this anisotropy model in the fitting.  This is the $\beta$-MBM
case. Finally, in the $\beta$-HM case, we adopt an
anisotropy profile that varies linearly with the logarithmic slope of
the number density profile:
\begin{equation}
\beta(r) = a + b\,{{\rm d}\ln \nu \over {\rm d}\ln r} \ ,
\label{betaHM}
\end{equation}
with $a=-0.15$ and $b=-0.19$, as was determined on a variety of
simulations by \cite{HM06}. Note that \citet{SH12} have argued that
cosmological haloes are better described by this relation than by the
attractor of \citet*{HJS10}, although \citet{Lemze+12} find that the
validity of this relation is limited to the central regions of haloes.

By fixing the anisotropy to determine the parameters of the mass profile,
these cases are similar in spirit to mass inversion \citep{MB10,Wolf+10},
except that the former requires a parametric form for the mass profile, while
the latter suffers from binning and extrapolation.

So, while the anisotropy parameter is
fixed in the $\beta$-iso case, it changes according to $r_{\rho}$ (a
free parameter here) in the $\beta$-MBM case, and according to the
logarithmic slope of the tracer number density profile (which is related to the
free parameter $r_{\nu}$) in the $\beta$-HM case.

\begin{table*}
\caption{MAMPOSSt results for 4 free parameters as well as several cases of
  constrained parameters}
\label{t:halores}
\tabcolsep=4pt
\begin{tabular}{lcrrrcrrrcrrrcrrr}
\hline
\hline
Case & $\beta(r)$ & \multicolumn{3}{c}{$r_{200}$} & &
\multicolumn{3}{c}{$r_{\nu}$} &  & \multicolumn{3}{c}{$r_{\rho}$} & & \multicolumn{3}{c}{anisotropy} \\
\cline{3-5}
\cline{7-9}
\cline{11-13}
\cline{15-17}
& & bias & ineff. & corr. & & bias & ineff. & corr. & & bias & ineff. &
corr. & & bias & ineff. & corr. \\
\hline
  Gen &  Cst &   0.004 & 0.040 &   {\bf 0.909} & & 0.027 & 0.102 &   {\bf 0.835} & &  0.032 & 0.217 &   {\bf 0.578} & &   0.007  & 0.073 &    --0.255  \\
  Gen & ML   & --0.003 & 0.040 &   {\bf 0.904} & & 0.024 & 0.104 &   {\bf 0.832} & & 0.057 & 0.229 &   {\bf 0.601} & & --0.221  & 0.887 &    --0.172  \\
  Gen &  T   & --0.006 & 0.040 &   {\bf 0.903} & & 0.026 & 0.103 &   {\bf 0.838} & & 0.039 & 0.169 &   {\bf 0.709} & &  0.007  & 0.085 & {\bf 0.621} \\
\hline
Split &  Cst &   0.004 & 0.040 &   {\bf 0.909} & & \em{0.032} & \em{0.101}  & {\bf \em   0.833} & &  0.036 & 0.216 &  {\bf 0.580} & &   0.010  & 0.074 &    --0.277  \\
Split & ML   & --0.003 & 0.040 &   {\bf 0.899} & & \em{0.032} & \em{0.101}  & {\bf \em   0.833} & &  0.044 & 0.218 &  {\bf 0.591} & & --0.072  & 0.756 &      0.038  \\
Split &  T   & --0.006 & 0.040 &   {\bf 0.907} & & \em{0.032} & \em{0.101}  & {\bf \em   0.833} & &  0.038 & 0.173 &  {\bf 0.713} & &   0.003  & 0.083 & {\bf 0.627} \\
\hline
 KVir &  Cst &\multicolumn{1}{c}{---}&\multicolumn{1}{c}{---}&\multicolumn{1}{c}{---}& & \em{0.021} & \em{0.103}  & {\bf \em  0.817} & &  0.017 & 0.305 &      0.266  & &   0.004 & 0.084 &      0.016  \\
 KVir & ML   &\multicolumn{1}{c}{---}&\multicolumn{1}{c}{---}&\multicolumn{1}{c}{---}& & \em{0.021} & \em{0.103}  & {\bf \em  0.817} & &  0.044 & 0.244 & {\bf 0.445} & & --0.237 & 1.564 &      0.221  \\
 KVir &  T   &\multicolumn{1}{c}{---}&\multicolumn{1}{c}{---}&\multicolumn{1}{c}{---}& & \em{0.021} & \em{0.103}  & {\bf \em  0.817} & &  0.040 & 0.198 & {\bf 0.638} & & --0.036 & 0.143 &      0.198  \\
\hline
 EVir &  Cst & \em{--0.000} & \em{0.042} &  {\bf \em  0.884} & & \em{0.028} & \em{0.103}  & {\bf \em  0.832} & &  0.041 & 0.223 & {\bf  0.596} & &   0.019 & 0.064 &     --0.155  \\
 EVir &   ML & \em{--0.000} & \em{0.042} &  {\bf \em  0.884} & & \em{0.028} & \em{0.103}  & {\bf \em  0.832} & &  0.071 & 0.201 & {\bf  0.571} & & --0.314 & 0.827 &       0.248  \\
 EVir &    T & \em{--0.000} & \em{0.042} &  {\bf \em  0.884} & & \em{0.028} & \em{0.103}  & {\bf \em  0.832} & &  0.028 & 0.185 & {\bf  0.714} & & --0.018 & 0.086 &       0.431  \\
\hline
 $\Lambda$CDM &  Cst &   0.008 & 0.038 &  {\bf 0.906} & & \em{0.032} & \em{0.101}  & {\bf \em  0.833} & & \em{0.070} & \em{0.151} & {\bf \em  0.674} & &   0.024 & 0.066 &   {\bf 0.470} \\
 $\Lambda$CDM & ML   &   0.002 & 0.039 &  {\bf 0.911} & & \em{0.032} & \em{0.101}  & {\bf \em  0.833} & & \em{0.061} & \em{0.154} & {\bf \em  0.678} & & --0.048 & 0.797 &   {\bf 0.493} \\
 $\Lambda$CDM &  T   & --0.004 & 0.039 &  {\bf 0.903} & & \em{0.032} & \em{0.101}  & {\bf \em  0.833} & & \em{0.061} & \em{0.147} & {\bf \em  0.675} & &   0.040 & 0.190 &   {\bf 0.497} \\
\hline
  MfL &  Cst &  0.005  & 0.038 &   {\bf 0.901} & & \em{0.032} & \em{0.101}  & {\bf \em  0.833} & & \multicolumn{1}{c}{---}&\multicolumn{1}{c}{---}&\multicolumn{1}{c}{---} & & 0.002 & 0.081 &  0.127  \\
  MfL & ML   & --0.001 & 0.039 &   {\bf 0.899} & & \em{0.032} & \em{0.101}  & {\bf \em  0.833} & & \multicolumn{1}{c}{---}&\multicolumn{1}{c}{---}&\multicolumn{1}{c}{---} & & 0.122 & 0.906 &  0.182  \\
  MfL &  T   & --0.003 & 0.041 &   {\bf 0.887} & & \em{0.032} & \em{0.101}  & {\bf \em  0.833} & & \multicolumn{1}{c}{---}&\multicolumn{1}{c}{---}&\multicolumn{1}{c}{---} & & 0.011 & 0.193 &  0.194  \\
\hline
  TLM & Cst &   0.003 & 0.040 &   {\bf 0.908} & & \multicolumn{1}{c}{---}&\multicolumn{1}{c}{---}&\multicolumn{1}{c}{---} & & --0.035 & 0.276 & {\bf 0.440} & & --0.008 & 0.072 &    --0.136  \\
  TLM & ML  & --0.006 & 0.046 &   {\bf 0.889} & & \multicolumn{1}{c}{---}&\multicolumn{1}{c}{---}&\multicolumn{1}{c}{---} & & --0.009 & 0.291 & {\bf 0.422} & &   0.116 & 0.715 &      0.183  \\
  TLM &  T  & --0.003 & 0.042 &   {\bf 0.906} & & \multicolumn{1}{c}{---}&\multicolumn{1}{c}{---}&\multicolumn{1}{c}{---} & &   0.020 & 0.213 & {\bf 0.605} & & --0.024 & 0.105 & {\bf 0.529} \\
\hline
  $\beta$-iso & Cst &   0.003 & 0.038 & {\bf 0.914} & & \em{0.032} & \em{0.101} & {\bf \em  0.833} & & --0.097 & 0.215 & {\bf 0.642} & & \em{--0.081} & \em{0.045} &\multicolumn{1}{c}{---}\\
 $\beta$-MBM  & ML  & --0.005 & 0.040 & {\bf 0.906} & & \em{0.032} & \em{0.101} & {\bf \em  0.833} & &   0.024 & 0.196 & {\bf 0.651} & &   \em{0.044} & \em{0.455} & {\bf \em  0.472} \\
 $\beta$-HM  & $\beta(\rho)$  & --0.003 & 0.040 & {\bf 0.899} & & \em{0.032} & \em{0.101} & {\bf \em  0.833} & &  0.009  & 0.222 & {\bf 0.617} & & \multicolumn{1}{c}{---} & \multicolumn{1}{c}{---}  &  \multicolumn{1}{c}{---}   \\
\hline
\end{tabular}

\parbox{\hsize}{Notes: 
These results are for 11 haloes of 500 particles, each observed along 3 axes
  out to the true value of $r_{200}$ in projection, with NFW density models
  and the `Clean' interloper removal algorithm.
Col.~1: Case for MAMPOSSt analysis
(see Table~\ref{t:cases});
col.~2: anisotropy model (Cst: $\beta=\rm cst$; ML:
  eq.~[\ref{e:ML}], \citealp{ML05b}; T: eq.~[\ref{e:T}], adapted from \citealp{Tiret+07});
cols.~3--5: virial radius;
cols.~6--8: tracer scale radius;
cols.~9--11: dark matter scale radius;
cols.~12--14: velocity anisotropy (i.e., ${\cal A}$ for the Cst model, $r_{\beta}$
for the ML model, and ${\cal A}_{\infty}$ for the T model).
The columns `bias' and `ineff.' respectively provide the mean and standard deviation (both computed with the biweight  technique) of $\log (o/t)$, while columns `corr.' list the
Spearman rank correlation coefficients between the true values and  MAMPOSSt-recovered ones
(values
in boldface indicate significant correlations between $o$ and $t$
values at the $\geq 0.99$ confidence level).
Values in italics indicate parameters that are not free in the MAMPOSSt
analysis for the Case considered.}
\end{table*}

\begin{figure}
\includegraphics[width=\hsize]{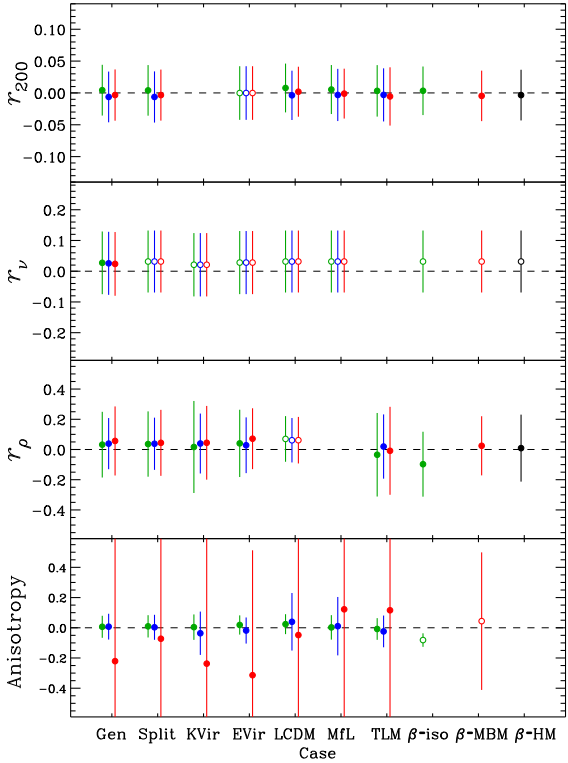}
\caption{Same as Fig.~\ref{f:gen} for the different cases of parameter
  constraints (see text and
  Table~\ref{t:cases}).
As in Fig.~\ref{f:gen}, results for the
  anisotropy models Cst, T, and ML are shown, left to right, in \emph{green},
\emph{blue}, and \emph{red}, respectively, and here the results for the 
$\beta$-HM anisotropy
model are shown in \emph{black}.  \emph{Open symbols}
  refer to those parameters that are not free parameters in the
  MAMPOSSt analysis for the Case considered. 
}
\label{f:halores}
\end{figure}

In Table~\ref{t:halores}, we list the results of our analysis for the
different observational cases for the 500-particle samples.  These
results are graphically displayed in Fig.~\ref{f:halores}. For the sake
of comparison, we list and plot again here the results for the Gen case,
already displayed in Table~\ref{t:gen} and Fig.~\ref{f:gen}.

The results for $r_{200}$ are essentially independent of the case considered.
While the  $r_{200}$ values for the EVir case are directly obtained  from the
Clean interloper removal scheme (App.~\ref{clean}), they are measured with
similar bias and inefficiency as for the Gen case.
In other words, $\sigma_z$ can be used to provide an estimate of 
$r_{200}$ with a comparable accuracy to that provided by MAMPOSSt.
The advantage of MAMPOSSt is of course that it provides estimates for
the other dynamical parameters at the same time.

Fitting $r_{\nu}$ from the $R$ distribution only, i.e. externally from
MAMPOSSt, does not significantly alter the accuracy returned for this parameter.
This means that incompleteness in the sample of tracers, if
properly accounted for, is not a significant issue for MAMPOSSt. Moreover,
this lifts our concern that the Split method does not find the same minimum
for $-\ln {\cal L}$ as the Gen (joint) method (see end of Sect.~\ref{genmeth}).
Using $r_{\rho}$ to predict $r_{\nu}$ (the TLM case) leads to larger bias and
inefficiency on $r_{\nu}$, but this occurs because the bias and inefficiency 
on $r_{\rho}$ are worse than those on
$r_{\nu}$.

Also, the results for $r_{\rho}$ depend little on the case
considered. The best results are obtained for the $\Lambda$CDM case, where
$r_{\rho}$ is not a free parameter of the MAMPOSSt analysis, but it is
estimated from $r_{200}$ using a theoretical relation. The good
results obtained for the $\Lambda$CDM case are not surprising, given that
$r_{200}$ is better constrained than $r_{\rho}$ in MAMPOSSt, and that
the test haloes considered here are extracted from a $\Lambda$CDM
cosmological simulation, and have therefore similar properties to
those used by \citet{MDvdB08} to establish the mass-concentration (and
hence the $r_{200}-r_{\rho}$) relation.

As for the anisotropy parameters, the ML model remains impossible to
constrain in all cases. For the Cst and T models, the bias and
inefficiency do not depend strongly on the case considered. However,
the $o$ vs. $t$ correlation values do depend on the case: a
significant correlation is obtained for ${\cal A}_{\infty}$ for the
Gen, Split, $\Lambda$CDM and TLM cases, but not for KVir, EVir, or
MfL.
For the $\Lambda$CDM and $\beta$-MBM cases, a significant correlation
is found even for the poorly constrained $r_{\beta}$ parameter, but
note that in the $\beta$-MBM case, $r_{\beta}$ is not a free parameter
of the MAMPOSSt analysis, but is forced to match $r_{\rho}$.

\begin{figure}
\centering
\includegraphics[width=\hsize]{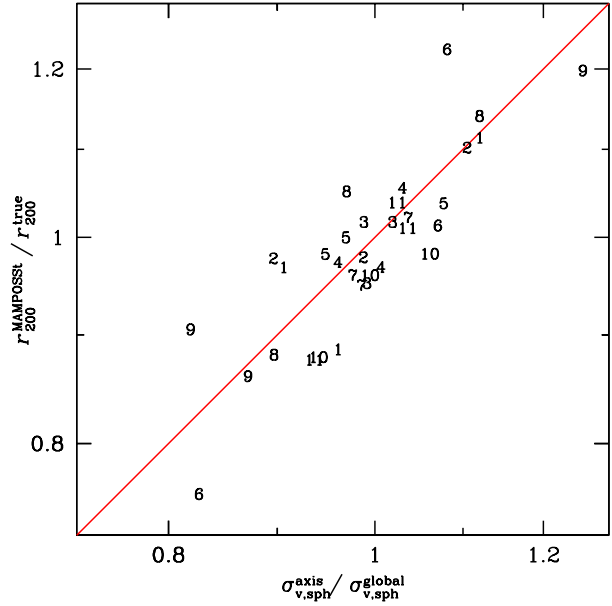}
\caption{Ratio of MAMPOSSt-estimated $r_{200}$ (Gen, NFW, T anisotropy) over
  true value versus the ratio of velocity dispersions within the true virial
  sphere, measured along the projection axis over that measured globally, for
  the $3\times11$ haloes.
  The \emph{numbers} indicate the halo rank following the order of increasing
  rank with increasing true $r_{200}$ of Table~\ref{t:haloprops}.  }
\label{f:svxyz}
\end{figure}

It is surprising that MAMPOSSt obtains worse inefficiencies in the KVir case
where $r_{200}$ is assumed perfectly known in comparison to the Gen, Split or
EVir cases.
The reason for this is probably related to the halo
triaxialities. When we observe a halo along a given line-of-sight, we
are sampling its velocity distribution only along that line-of-sight.
If the components of the halo velocity distribution along different
axes are different, this difference will be reflected in our results.

This point is demonstrated in Fig.~\ref{f:svxyz}. There, we show the
correlation between the ratio of $r_{200}$ measured by MAMPOSSt to the true
value measured in 3D versus the ratio of 
velocity dispersions within the true virial sphere, measured along the
projection axis over that measured globally.
The very high correlation (Spearman rank: $r=0.82$) shows that the error on
$r_{200}$ is related to the velocity dispersion ratio above, which directly
measures the triaxiality of the halo (without mixing with the effects of
interlopers beyond the virial sphere).

\subsection{Stacked haloes}
\label{s:stacks}
To extract all the possible information from the available data, it is
a common practice to construct stacked samples
\citep[e.g.][]{CYE97,BG03,KBM04,BP09}.  In the case of clusters of
galaxies, this is done by scaling the galaxy distances from their
cluster centres and the galaxy velocities with respect to their
cluster mean velocity, by their cluster $r_{200}$ and $v_{200}$,
respectively, where $v_{200} \equiv (G M_{200}/r_{200})^{1/2}$.  To do
this, prior knowledge of the individual cluster $r_{200}$ values is
needed.  Trying to mimic the observational situation as close as
possible, we stack our haloes using the values of 
$r_{200}$ directly obtained from the Clean interloper removal scheme, as
described in Appendix~\ref{clean}, i.e. the values used in what we
called the EVir case in Sect.~\ref{s:testothers}.  We thus build three
stacked haloes, one for each projection axis, from the eleven 500-particle
haloes, each first passed through the Clean interloper removal scheme
(app.~\ref{clean}). 

The stacks thus created from
the 500-particle (respectively 100-particle) samples contain $5248,
5260, 5213$ (respectively $1061, 1065, 1048$) DM particles along the
$x$-, $y$-, and $z$-axis, respectively.

\begin{table}
\begin{center}
\caption{MAMPOSSt results for the stacked haloes}
\label{t:stack}
\tabcolsep=3pt
\begin{tabular}{ccrrrrcrr}
\hline
\hline
Projection & $\beta(r)$ & $r_{200}$ & $r_{\nu}$  &
\multicolumn{2}{c}{$r_{\rho}$} & & 
\multicolumn{2}{c}{anisotropy} \\
\cline{5-6}
\cline{8-9}
 & & & & stack & mean & & stack & mean \\
\hline
\hline
\multicolumn{9}{c}{$11\times500$} \\
\hline
$x$ & Cst &  \em{1.104} & \em{0.208} &  0.246 & \em{0.227} & & 1.195 & \em{1.230} \\ 
$y$ & Cst &  \em{1.050} & \em{0.186} &  0.199 & \em{0.292} & & 1.184 & \em{1.253}  \\
$z$ & Cst &  \em{1.018} & \em{0.136} &  0.146 & \em{0.260} & & 1.130 & \em{1.240}  \\
\multicolumn{2}{l}{bias}  & \em{--0.014} &  \em{--0.166} & --0.129 & \em{--0.003} & & --0.008 & \em{0.011} \\
\multicolumn{2}{l}{inefficiency} &   \em{0.021} &    \em{0.109} &   0.134 &   \em{0.065} & &   0.014 & \em{0.005} \\
\hline
$x$ & ML  &  \em{1.104} & \em{0.208} &  0.219 & \em{0.328} & & 0.801 & \em{0.045}  \\
$y$ & ML  &  \em{1.050} & \em{0.186} &  0.203 & \em{0.306} & & 0.429 & \em{0.049}  \\
$z$ & ML  &  \em{1.018} & \em{0.136} &  0.154 & \em{0.259} & & 0.494 & \em{0.256}  \\
\multicolumn{2}{l}{bias}  & \em{--0.014} &  \em{--0.166} & --0.120 & \em{  0.060} & &  0.305 & \em{--0.743} \\
\multicolumn{2}{l}{inefficiency} &   \em{0.021} &    \em{0.109} &   0.090 &   \em{0.061} & &  0.160 & \em{  0.446} \\
\hline
$x$ & T   &  \em{1.104} &  \em{0.208} &  0.283 & \em{0.337} & & 1.565 & \em{1.330}  \\
$y$ & T   &  \em{1.050} &  \em{0.186} &  0.235 & \em{0.256} & & 1.509 & \em{1.330}  \\
$z$ & T   &  \em{1.018} &  \em{0.136} &  0.180 & \em{0.252} & & 1.459 & \em{1.427}  \\
\multicolumn{2}{l}{bias}  & \em{--0.014} &  \em{--0.166} & --0.054 & \em{--0.010} & &  0.018 & \em{--0.038} \\
\multicolumn{2}{l}{inefficiency} &   \em{0.021} &    \em{0.109} &   0.116 &   \em{0.075} & &  0.018 & \em{  0.018} \\
\hline
\multicolumn{9}{c}{$11\times100$} \\
\hline
$x$ & Cst &  \em{1.107} & \em{0.150} &  0.143 & \em{0.160} & & 1.101 & \em{1.164}  \\
$y$ & Cst &  \em{1.016} & \em{0.190} &  0.144 & \em{0.171} & & 1.224 & \em{1.402}  \\
$z$ & Cst &  \em{1.038} &  \em{0.142} &  0.146 & \em{0.210} & & 1.151 & \em{1.181}  \\
\multicolumn{2}{l}{bias}  & \em{--0.017} &  \em{--0.238} & --0.256 & \em{--0.169} & & --0.019 & \em{--0.014} \\
\multicolumn{2}{l}{inefficiency} &   \em{0.022} &    \em{0.075} &   0.050 &   \em{0.070} & &  0.027 & \em{0.048} \\
\hline
$x$ & ML  &  \em{1.107} &  \em{0.150} &  0.155 & \em{0.178} & & 0.515 & \em{0.118}  \\
$y$ & ML  &  \em{1.016} &  \em{0.190} &  0.168 & \em{0.212} & & 0.093 & \em{0.051}  \\
$z$ & ML  &  \em{1.038} &  \em{0.142} &  0.163 & \em{0.228} & & 0.263 & \em{0.226}  \\
\multicolumn{2}{l}{bias}  & \em{--0.017} &  \em{--0.238} & --0.205 & \em{--0.100} & & --0.043 & \em{--0.368} \\
\multicolumn{2}{l}{inefficiency} &   \em{0.022} &    \em{0.075} &   0.021 &   \em{0.064} & &  0.439 & \em{  0.382} \\
\hline
$x$ & T   &  \em{1.107} &  \em{0.150} &  0.189 & \em{0.183} & & 1.515 & \em{1.506}  \\
$y$ & T   & \em{1.016} &  \em{0.190} &  0.164 & \em{0.202} & & 1.464 & \em{1.424}  \\
$z$ & T   & \em{1.038} &  \em{0.142} &  0.183 & \em{1.424} & & 1.554 & \em{1.276}  \\
\multicolumn{2}{l}{bias}  & \em{--0.017} &  \em{--0.238} & --0.158 & \em{--0.130} & &   0.018 & \em{--0.014} \\
\multicolumn{2}{l}{inefficiency} &   \em{0.022} &    \em{0.075} &   0.036 &   \em{0.025} & &  0.015 & \em{  0.042} \\
\hline
\end{tabular}
\end{center}

\parbox{\hsize}{Notes: 
Col~1: viewing axis;
col.~2: anisotropy model;
col.~3: virial radius (biweight mean over 11 haloes from Clean interloper
removal scheme);
col.~4: tracer scale radius (biweight mean of MAMPOSSt EVir estimate for the
11 haloes);
col.~5: DM scale radius (MAMPOSSt);
col.~6: same (biweight mean of MAMPOSSt EVir estimate for the 11 haloes);
col.~7: anisotropy parameter (${\cal A}$ for Cst, $r_\beta$ for ML and ${\cal
  A_\infty}$ for T) from MAMPOSSt;
col.~8: same (biweight mean of MAMPOSSt EVir estimate for the 11 haloes).
Values  that are
not directly obtained through MAMPOSSt on the stacked halo but from the mean
of the individual haloes are shown in \emph{italics}. 
The biases and inefficiencies are respectively  computed from biweight means
and gapper standard deviations  of the 3 values (see \citealp{BFG90}). 
}
\end{table}

\begin{figure}
\includegraphics[width=\hsize]{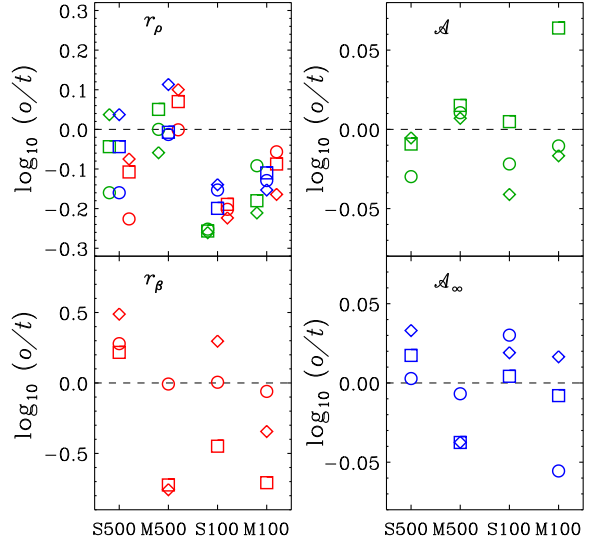}
\caption{MAMPOSSt residuals for stacked haloes (labelled `S500' and `S100',
  when 
 built from the 11 haloes, respectively sampled with 500 and 100 particles),
and biweight mean of the values obtained for
  the 11 haloes in the EVir case (see Sect.~\ref{s:testothers},
  labelled `M500' and `M100', for the 500 and 100 particle haloes,
  respectively). 
  Different
  symbols identify different projections, $x$-axis: \emph{diamonds}, $y$-axis:
  \emph{squares}, $z$-axis: \emph{circles}.  In the \emph{upper-left panel} ($r_{\rho}$), the
  \emph{green}, \emph{blue}, and \emph{red symbols} are the results obtained using,
  respectively, the Cst, T, and ML anisotropy models.}
\label{f:stackres}
\end{figure}

Since we need to fix the $r_{200}$ values of the haloes before
stacking, $r_{200}$ cannot be a free parameter of the MAMPOSSt
analysis. Moreover, in an effort to mimick the observational
procedure, we estimate the individual $r_{\nu}$ values of the 11 haloes
also before stacking, and then we take the biweight average of these
values as representative of the $r_{\nu}$ of the stack. In the real
world, this is done because individual haloes often suffer from
different spectroscopic incompleteness levels. More precisely, we take
the $r_{\nu}$ estimates obtained in the EVir case (see
Sect.~\ref{s:testothers}).  The remaining two parameters, $r_{\rho}$
and the anisotropy parameter are estimated via the MAMPOSSt analysis.
Results for the six stacked haloes are listed in Table~\ref{t:stack}
and displayed in Fig.~\ref{f:stackres}. 
We use the gapper estimate of dispersion (\citealp{WT76}, see
\citealp{BFG90}) given our small sample ($N=3$). 

\begin{figure}
\includegraphics[width=\hsize]{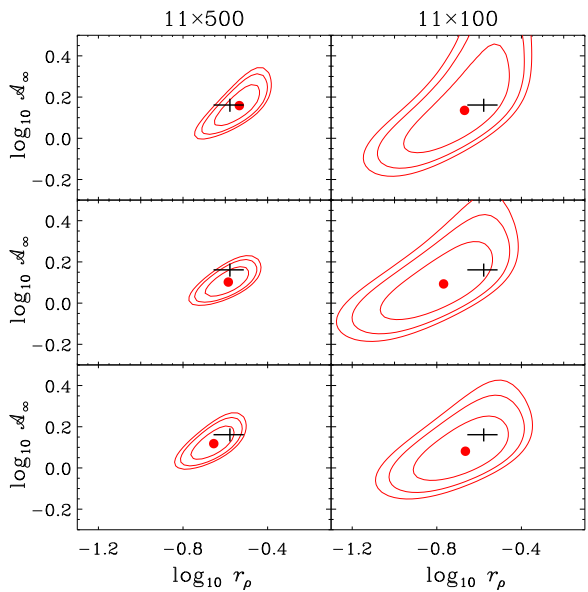}
\caption{MAMPOSSt confidence contours in the $\log r_{\rho} - \log
  A_{\infty}$ plane for the stacked haloes constructed from single-axis
  projections of 11 haloes from the 500-particle (left panels) and
  100-particle (right panels) samples. The $x$-axis, $y$-axis, and $z$-axis
  stacks are shown in the \emph{top}, \emph{middle}, and \emph{bottom panels},
  respectively.  The contours are drawn at 1-, 2- and 3-$\sigma$. The
  best-fit solutions are indicated by the \emph{filled circles}, while the 
\emph{crosses}
  represent the biweight average of the true values of the parameters
  for the 11 haloes. The cross lengths across the $x$- and $y$-axes
  represent the errors on the two parameter averages (see
  Table~\ref{t:haloprops}).  }
\label{f:stack}
\end{figure}
In Fig.~\ref{f:stack}, we show
1-, 2-, and 3-$\sigma$ contours in the $\log
r_{\rho} - \log {\cal A}_{\infty}$ plane (for the T anisotropy model). In this
figure, we indicate with a cross the expected solution for the stacked
haloes.  This is the biweight-average (with its error) of the true
$r_{\rho}$ and ${\cal A}_{\infty}$ values of the 11 individual 
haloes from which
the stacked haloes are constructed. These values are also listed in the
last line of Table~\ref{t:haloprops}. The average values of $r_{-2}$
and $r_{\beta}$ are identical, in substantial agreement with what was
found by {MBM10}. Note also that the average value of
$A_{\infty}$ corresponds to $\beta_{\infty}=0.5$, meaning that the
average ML and T anisotropy models are the same.

Fig.~\ref{f:stack} shows that the expected solution for the
stack sample is always within the 1-$\sigma$ contour of the MAMPOSSt
result for the T model. While not shown, this is also the case for the
Cst model, but not for the ML model, where
$r_{\beta}$ remains essentially
unconstrained, as was the case for the individual haloes. The estimates of the
parameter $r_{\rho}$ do not depend 
on the assumed anisotropy model. They are biased low for the stacks
built from the 100-particle samples, even if not significantly so (see
Fig.~\ref{f:stack}). Interestingly, while the bias is stronger for the
100-particle stack than for the 500-particle one, the dispersion of
the recovered values appears to be lower. The dispersion is instead
higher, as expected, for the anisotropy parameters, except for ${\cal
  A}_{\infty}$ where the dispersion is comparable for the 500-particle
and the 100-particle stacks. However, with only three results per
anisotropy model for each Stack sample, these differences in bias and
dispersions do not appear to be statistically significant.

In Table~\ref{t:stack}, we also list 
the biweight average values of the 11 individual haloes along
each projection, obtained in the EVir case (see
Sect.~\ref{s:testothers}), and these values are also plotted in
Fig.~\ref{f:stackres} (labelled `M') together with the best-fit solutions for the
stacked haloes (labelled `S'). Comparing the best-fit values of $r_{\rho}$ and
the anisotropy parameters obtained from the stacked haloes with those found from the mean
of the individual 11 haloes (along the same projection), it is difficult to
decide whether it is better to run MAMPOSSt on the stacked halo or to adopt
means of the individual MAMPOSSt estimates. 
A close inspection of Table~\ref{t:stack} and
Fig,~\ref{f:stackres} indicates that, for $N=500$, stacks are more accurate for estimating
$r_\beta$ and ${\cal A}_\infty$, while means of
individual MAMPOSSt estimates appear more accurate for estimating $r_\rho$
and ${\cal A}$.
But the statistics are poor.
In the real observational situation, it therefore seems
advisable to consider both the average of the results of the individual
haloes and the result obtained for the stacked sample.

\subsection{A large halo}

\begin{table}
\begin{center}
\caption{MAMPOSSt results for the most massive halo, 5726}
\label{t:5726}
\tabcolsep=4pt
\begin{tabular}{crcrrrr}
\hline
\hline
Projection & \multicolumn{1}{c}{$N$} & $\beta(r)$ & $r_{200}$ & $r_{\nu}$  & $r_{\rho}$ & anisotropy \\
\hline
$x$ & 4529 & Cst &  1.538 & 0.499 &  0.262 & 1.053 \\ 
$y$ & 4520 & Cst &  1.747 & 0.369 &  0.352 & 1.122  \\
$z$ & 4662 & Cst &  1.642 & 0.318 &  0.197 & 1.252  \\
\multicolumn{2}{l}{bias} &  & --0.005 &  --0.024 & --0.190 & --0.096 \\
\multicolumn{2}{l}{inefficiency} & &   0.033 &    0.116 &   0.149 &   0.044 \\
\hline
$x$ & 4529 & ML  &  1.505 & 0.499 &  0.308 & 0.568 \\ 
$y$ & 4520 & ML  &  1.711 & 0.370 &  0.352 & 0.638  \\
$z$ & 4662 & ML  &  1.621 & 0.318 &  0.190 & 0.301  \\
\multicolumn{2}{l}{bias} &  & --0.013 &  --0.023 & --0.142 &   1.080 \\
\multicolumn{2}{l}{inefficiency} & &   0.033 &    0.115 &   0.158 &   0.192 \\
\hline
$x$ & 4529 & T   &  1.454 & 0.502 &  0.447 & 2.289 \\ 
$y$ & 4520 & T   &  1.700 & 0.371 &  0.425 & 1.666  \\
$z$ & 4662 & T   &  1.623 & 0.320 &  0.236 & 1.705  \\
\multicolumn{2}{l}{bias} &  & --0.017 &  --0.021 &   0.030 & --0.116 \\
\multicolumn{2}{l}{inefficiency} & &   0.040 &    0.115 &   0.164 &   0.082 \\
\hline
\hline
$x$ & 489 & Cst &  1.616 & 0.492 &  0.336 & 0.887 \\ 
$y$ & 460 & Cst &  1.773 & 0.398 &  0.340 & 1.156  \\
$z$ & 476 & Cst &  1.688 & 0.260 &  0.187 & 1.188  \\
\multicolumn{2}{l}{bias} &  &   0.008 &  --0.036 & --0.081 & --0.084 \\
\multicolumn{2}{l}{inefficiency} & &   0.024 &    0.164 &   0.154 &   0.075 \\
\hline
$x$ & 489 & ML  &  1.554 & 0.494 &  0.483 & 0.764 \\ 
$y$ & 460 & ML  &  1.739 & 0.399 &  0.335 & 0.562  \\
$z$ & 476 & ML  &  1.672 & 0.261 &  0.195 & 0.310  \\
\multicolumn{2}{l}{bias} &  & --0.001 &  --0.034 & --0.107 &   1.015 \\
\multicolumn{2}{l}{inefficiency} & &   0.029 &    0.164 &   0.232 &   0.231 \\
\hline
$x$ & 489 & T   &  1.454 & 0.496 &  0.676 & 3.454 \\ 
$y$ & 460 & T   &  1.724 & 0.401 &  0.412 & 1.735  \\
$z$ & 476 & T   &  1.677 & 0.261 &  0.247 & 1.829  \\
\multicolumn{2}{l}{bias} &  &   0.010 &  --0.033 &   0.003 & --0.092 \\
\multicolumn{2}{l}{inefficiency} & &   0.044 &    0.165 &   0.258 &   0.177 \\
\hline
\end{tabular}
\end{center}

\vbox{Notes: 
The results are for Case Gen.
Col.~1: viewing axis;
col.~2: number of particles after interloper removal with the Clean method;
col.~3: anisotropy model;
col.~4: virial radius;
col.~5: tracer scale radius;
col.~6: DM scale radius;
col.~7: anisotropy parameter (${\cal A}$ for Cst, $r_\beta$ for ML and ${\cal
  A_\infty}$ for T);
The biases and inefficiencies are respectively  computed from biweight means
and gapper standard deviations  of the 3 values (following \citealp{BFG90}). 
}
\end{table}

It is interesting to consider whether we can achieve the same accuracy
using a single halo, but with a total number of tracer particles similar
to that of the stack. We extract $\sim 5000$ particles within $r_{200}$
for the more massive halo in our sample, 5726, remove interlopers with
the Clean method, and run MAMPOSSt for the general case, adopting
the NFW mass density and different anisotropy models. 

Table~\ref{t:5726} compares the results for the 3 cones of 500 particles
(counted before interloper removal) and the 3 with $\simeq$5000 particles.
We use again the gapper  dispersion, given our small sample ($N=3$). 
As expected, the biases and inefficiencies are generally reduced when
selecting all ($\simeq$5000) particles instead of randomly selecting 500.
For the T anisotropy model, the inefficiencies in $\log(o/t)$ are reduced
by 0.1 dex for $r_\rho$ and anisotropy, 0.15 dex for $r_\nu$, but virtually
unchanged for $r_{200}$.
Surprisingly, for Cst and ML anisotropy,
$r_{200}$ is reached with worse 
inefficiency with the full 5000-particle sample in comparison with the
500-particle one.

However, one should not over-interpret these comparisons.
Indeed, our Monte-Carlo tests on 10$\,$000 random samples of 3 objects
arising from a Gaussian distribution indicate that while the
gapper standard deviation is unbiased even for as few as 3 objects, it has an
inefficiency of 0.53 times its true value for $N=3$. In other words, an inefficiency of
$\log(o/t)$ of 0.1 would have roughly 0.05 accuracy. Therefore, any
improvement or worsening of inefficiency in $\log(o/t)$ of less than a factor
of 2 is clearly not statistically significant.
Nevertheless, the fact that most inefficiencies are reduced when increasing
the sample size from 500 to 5000 is much more significant, and is
indeed what is qualitatively expected. However, the expected quantitative improvement
of $\log \sqrt{10}=0.5$ does not appear to be attained for the inefficiency
of $\log(o/t)$.

The bias
and inefficiencies for the 5000-particle sample are typically higher for the single
halo than for the $11\times500$ particle stack (with just 10\% more
particles, although the extraction there was for case EVir instead of Gen).
But again, the inefficiencies are within a factor 2, so these changes are
individually not statistically significant. But the combination of all of
them appears to be statistically significant.

The worse inefficiencies for the large halo in comparison with the stacked
halo of similar number of tracers 
can be due to either our assumption of Gaussian
velocity distributions (see Sect.~\ref{gaussian}) or to the
triaxiality of the halo velocity ellipsoids. 
The velocity distribution of the stack halo approaches a Gaussian 
by 
the central limit theorem,
so our Gaussian assumption is better verified in a stack sample
than in individual haloes.  As a matter of fact, a number of studies
have shown the 3D velocity distributions of individual haloes to deviate
from Gaussianity \citep{WLGM05,HMZS06}.
But perhaps the
dominant effect is that of triaxiality. As discussed at the end of
Sect.~\ref{s:testothers}, the results obtained by MAMPOSSt for $r_{200}$
are influenced by the choice of the projection axis, and in the
present case, the $r_{200}$ values obtained for 5726 along the $x$, $y$,
and $z$ projections when using 5000 particles, are ordered in the same
way as those obtained when using 500 particles.

\section{Conclusions and Discussion}
\label{discus}

We have presented the formalism for a new method, called MAMPOSSt, for the determination
of the mass and anisotropy profiles of spherical systems, assumed to be in
dynamical equilibrium, in particular, as stationary systems with no
streaming motions.
In MAMPOSSt, the distribution of tracers in projected phase space are fit by
the predicted distribution arising from assumed 3D radial profiles of the
tracer density, total
mass, and velocity anisotropy, as well from an assumed family of shapes for
the radial and tangential components of the 3D velocity distribution. 

MAMPOSSt has several important advantages over other methods for
mass/anisotropy modeling:
\renewcommand{\theenumi}{\arabic{enumi})}
\begin{enumerate}
\item 
MAMPOSSt does not involve binning the data, hence its results are independent of the
  choice of such radial bins, in contrast with methods based upon velocity moments, including
  anisotropy and mass inversions;
\item MAMPOSSt does not involve interpolations and extrapolations of binned
  radial profiles of the observables, again in contrast with
anisotropy and mass inversions;
\item MAMPOSSt does not require differentiation of the observed projected
  pressure, $\Sigma(R)\,\sigma_z^2(R)$, once more in 
contrast with anisotropy and mass
  inversions;
\item MAMPOSSt extracts accurate constraints on the velocity anisotropy, in
  contrast with methods that assume Gaussian LOS velocity distributions;
\item MAMPOSSt is very fast, as it involves a single integral (for popular $\beta(r)$
  profiles). Indeed,  the calculations for the best fit parameters and
  their marginal and correlated distributions through MCMC for a 500-tracer
  system (as displayed in Fig.~\ref{figmcmc}) require roughly 4 minutes of
  CPU time on a standard desktop personal computer. By contrast,
  distribution function methods involve triple integrals, and therefore take
  typically 1000 times longer to run, i.e. a few days for 500-tracer
  systems. Orbit modeling techniques are even slower and cannot properly probe parameter space.
\end{enumerate}

In this work, we have extensively tested MAMPOSSt, for the case of Gaussian 3D
velocity distributions, on a set of 11 cluster-mass
haloes extracted from cosmological simulations. The results of these
tests indicate that, for systems with 500 velocities, MAMPOSSt provides essentially unbiased estimates
of the relevant mass and velocity anisotropy profile parameters, with
inefficiencies of 10\% for the virial radius, 20\% for the constant or outer
velocity anisotropy,
27\% for the tracer scale radius, but as high as 48\% for the scale radius of
the DM.
However, MAMPOSSt seems unable to
set constraints on the radius of transition of a gently rising
anisotropy model such as ML.

We have noted that the results of MAMPOSSt are similar when we inserted the 
virial radius among the parameters to be jointly fit or when we derived the
virial radius using the new interloper removal scheme that we presented in
Appendix~\ref{clean}.
We found that this
interloper-removal algorithm performs as well or better than other
methods. 
\begin{figure}
\centering
\includegraphics[width=\hsize]{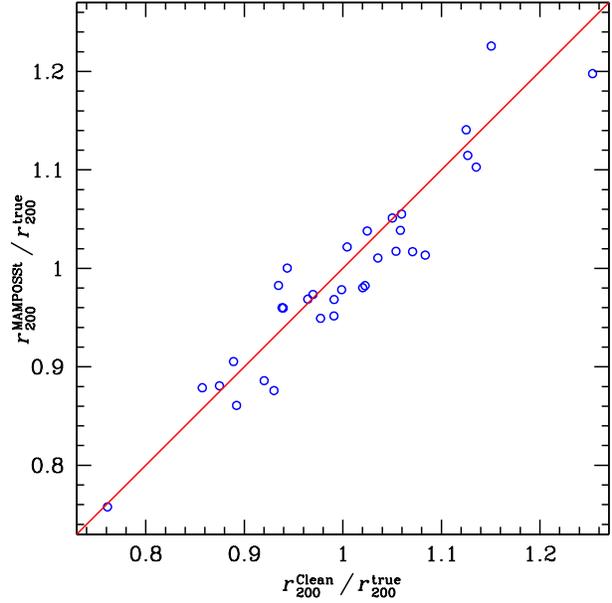}
\caption{Comparison of virial radius errors from two schemes: MAMPOSSt vs. Clean interloper
  removal scheme (appendix~\ref{clean}), for the $3\times11$ haloes. The
  \emph{red line} corresponds to equality.
\label{compare}}
\end{figure}
As shown in Fig.~\ref{compare}, the errors on $r_{200}$ from MAMPOSSt are highly
correlated with those obtained from the Clean procedure.  Hence, the
new interloper removal scheme produces extremely fast estimates of the
virial radius, based on the LOS velocity dispersion, that are very
close to those obtained with the full MAMPOSSt procedure. 

The
correlation highlighted in Fig.~\ref{compare} suggests that the quality of MAMPOSSt estimates is
limited by the accuracy of the interloper removal. However, we also
saw (Fig.~\ref{f:svxyz}) that the error on $\log r_{200}$ is correlated with the
ratio of velocity dispersions within the true virial sphere, measured
along the projection axis over that measured globally.  This
second correlation indicates that our capability of recovering the true
values of $r_{200}$ is limited by the effects of triaxiality, i.e. by the
fact that we only observe one component of the 3-dimensional velocity
distribution.  The effect of triaxiality may be even more important
than the uncertainties in the interloper cleaning process. 

Our additional tests lead to the following conclusions:
\begin{enumerate}
\item Best results are obtained setting all 4 parameters free and using the T
  anisotropy model.
\item Setting the virial radius to the true value (e.g. measured through
  other techniques, such as X-rays or lensing) leads to worse results on the
  other parameters than when the virial radius is an additional free
  parameter. We argue that this is caused by the triaxiality of the
  haloes
(Fig.~\ref{f:svxyz}).
\item MAMPOSSt works surprisingly well with samples of 100 tracers instead of
  500,  only somewhat better with samples of 5000 tracers, even better with
  stacked haloes of $11\times 500$ tracers, similar to the mean of these
  individual tracers, but not well as expected from $N^{-1/2}$ arguments.
\item MAMPOSSt is not very sensitive to the chosen aperture, when this is to
  within $\approx$35\% of the true value of $r_{200}$.
\end{enumerate}

\begin{table*}
\caption{Comparison of the biases and inefficiencies of MAMPOSSt with those
  of other methods}
\begin{center}
\begin{tabular}{lccrrcrrcrrcrr}
\hline
\hline
Method & Reference & Particles & 
\multicolumn{2}{c}{$r_{\rm v}$} & & 
\multicolumn{2}{c}{$r_\rho$} & & 
\multicolumn{2}{c}{${\cal A}$} & & 
\multicolumn{2}{c}{${\cal A}_\infty$} \\
\cline{4-5}
\cline{7-8}
\cline{10-11}
\cline{13-14}
 & & & 
bias & ineff. & & 
bias & ineff. & & 
bias & ineff. & & 
bias & ineff. \\
\hline
Dispersion-kurtosis & \cite{SLM04} & 400 & --0.023 & 0.033 & & 0.080 & 0.240 &
& --0.040 & 0.110 & & --- & --- \\
MAMPOSSt $\beta$=cst & this article & 400 & {\bf 0.004} & 0.040 & & {\bf --0.030} &
0.289 & & {\bf --0.010} & {\bf 0.069} & & --- & ---\\
Distribution function & \cite{WLMG09} & 300 & --0.026 & 0.048 & & 0.017 &
0.169 & & --- & --- & & --0.002 & 0.212 \\
MAMPOSSt $\beta$-T & this article & 300 & {\bf --0.003} & {\bf 0.044} & &
 0.017 & 0.198 & & --- & --- & &  --0.016 & {\bf 0.116} \\
\hline
\end{tabular}  
\end{center}

\parbox{\hsize}{Notes:
The biases and inefficiencies on virial masses for the dispersion-kurtosis
and distribution function methods were divided by 3 to convert to biases and
inefficiencies for virial radii.
All analyses are for case TLM, and for MAMPOSSt the Clean interloper removal
scheme is used.
Dispersion-kurtosis $\log(o/t)$ for the DM scale radius were taken from the
analogous values that \cite{SLM04} obtained for the concentration parameter
(basically assuming that in comparison, the errors on the virial radius are
small).
For the distribution function method, values of $\log (o/t)$ for the outer
anisotropy were measured at roughly the virial radius, while the MAMPOSSt
comparison (last row) is for the T anisotropy model (eq.~[\ref{e:T}]). 
The biases and inefficiencies are estimated with the biweight mean and
dispersion, except for the dispersion-kurtosis method, where they are taken from the
2nd line of Table 3 of \citeauthor{SLM04}.
Values where MAMPOSSt is better are highlighted in bold.
}
\label{comparemethods}
\end{table*}
Table~\ref{comparemethods} compares the accuracy of MAMPOSSt with that of the
dispersion-kurtosis method of \citeauthor{Lokas02} (\citeyear{Lokas02},
tested by \citealp*{SLM04}) and the DF
method of \cite{WLMG09}.
In comparison with the dispersion-kurtosis method, MAMPOSSt does slightly less well on
the inefficiencies of 
the virial radius and worse on that of the DM scale radius (but better on the
biases of both radii), 
but MAMPOSSt performs better for the constant anisotropy (both bias and inefficiency).
Moreover, MAMPOSSt has the advantage of being flexible enough to infer the outer
anisotropy, while the dispersion-kurtosis method of \cite{Lokas02} 
is limited to constant
anisotropy (but this limitation to the dispersion-kurtosis method has been
recently lifted by \citealp{RF12}).
Comparing with the much slower DF method, MAMPOSSt does 
somewhat worse on the inefficiency on the scale radius, with the same bias, 
yet MAMPOSSt is  respectively slightly (not significantly) and much better on the
inefficiencies on the  virial radius and 
outer anisotropy, with less bias on the former.
Note, however, that we forced here the inner velocities to be isotropic,
while these were free parameters in the DF method and that the DF method
computes the outer anisotropies at $5\,r_{\rm s}$ instead of the asymptotic
value.
Nevertheless, we found (Table~\ref{t:halores}) that MAMPOSSt always performs as well or  better when
all parameters are set free, and we would expect the same for the DF
method. Moreover, we see no reason why the inefficiency of the anisotropy
measured near the virial radius should be worse than that of its asymptotic value. 

Despite the very good results of our tests, it must be noted that
although the Gaussian MAMPOSSt recovers the 2nd velocity moment
(eq.~[\ref{siglosgaussian}]) , it
does not recover the fourth moment.  Indeed, taking the fourth moment
of the velocity distribution of equation~(\ref{pofvzgaussian}) leads
to
\begin{eqnarray}
\Sigma(R)\,\overline{v_z^4}(R) &\!\!\!\!=\!\!\!\!& 
\int_{-\infty}^{+\infty}  v_z^4 \,g(R,v_z)\,\d v_z \nonumber \\
&\!\!\!\!=\!\!\!\!& \sqrt{2\over\pi} 
\int_R^\infty {\nu(r)\over  \sigma_z(R,r)}\,{r\,\d r\over\sqrt{r^2-R^2}} \nonumber \\
&\!\!\!\!\mbox{}\!\!\!\!& \quad \times
\int_{-\infty}^{+\infty} v_z^4 \exp\left [-{v_z^2\over
    2\,\sigma_z^2(R,r)}\right]\,\d v_z \nonumber \\
&\!\!\!\!=\!\!\!\!&
6\,\int_R^\infty \nu \sigma_r^4 
\left[1-\beta(r) {R^2\over r^2}\right]^2\,{r\,\d r\over \sqrt{r^2-R^2}}
\ ,
\label{vlos4}
\end{eqnarray}
where the second equality of equation~(\ref{vlos4}) is obtained after
reversing the order of integration, while the final equality uses
equation~(\ref{sigmaz}).
Equation~(\ref{vlos4}) then gives the LOS velocity kurtosis excess
\begin{equation}
\kappa_z(R) = {\overline{v_z^4}(R)\over \sigma_z^4(R)}-3
\ .
\label{kappalos}
\end{equation}

Unfortunately, the expression for the fourth velocity moment in
equation~(\ref{vlos4}) differs from the expression found by \cite{Lokas02}
from the 4th order Jeans equation in the case of $\beta = \rm cst$.
At large projected radius ($R \approx r_{\rm v}$),
$\kappa_z$ estimated by equations~(\ref{vlos4}) and (\ref{kappalos}) 
turns out to be nearly independent of $\beta$.
This shows the limit of the Gaussian approximation for the 3D velocity distribution.

Nevertheless, despite the poor adequacy of the Gaussian approximation, which
does not produce the correct LOS kurtosis profiles, MAMPOSSt with
Gaussian velocities performs quite well according to our numerous tests.
We are therefore confident that the
method is mature for being used on real data-sets. And we are preparing
analyses on several scales: dwarf spheroidal galaxies, giant ellipticals
(traced by planetary nebulae), groups and clusters of galaxies.
We also
plan to test MAMPOSSt on elliptical galaxies formed in dissipative
mergers.

One should note that in $\Lambda$CDM
haloes, the 3D velocity distribution, measured in shells, is not Gaussian
\citep{WLGM05}, but resembles the
$q$-Gaussian distribution, which generalizes the Gaussian distribution in the same way as
\cite{Tsallis88} entropy generalizes Boltzmann-Gibbs entropy.
The $q$-Gaussian (often called Tsallis) fits well a host of velocity distributions found in
single-component dissipationless self-gravitating systems, with an index that
varies linearly with the slope of the density profile \citep{HMZS06}. 
We are thus
preparing a generalization of  MAMPOSSt to the Tsallis distribution of 3D
velocities (appropriately combining radial and tangential terms), and expect
this to perform even better than the MAMPOSSt-Gaussian algorithm.

\section*{Acknowledgements}

We gratefully acknowledge the anonymous referee for constructive comments
that strengthened our work, and 
Leandro Jos\'e Beraldo e Silva for his critical
reading of the manuscript. We thank Giuseppe Murante for providing us with the cosmological simulation and
Anthony Lewis for developing and maintaining the CosmoMC MCMC code.
Thanks also to Michael Powell for making his NEWUOA minimization code public,
and Radek Wojtak for supplying his test
output in digital form.

\bibliography{mamposst3}

\begin{thebibliography}{70}
\expandafter\ifx\csname natexlab\endcsname\relax\def\natexlab#1{#1}\fi

\bibitem[{{Ascasibar} \& {Gottl{\"o}ber}(2008)}]{AG08}
{Ascasibar} Y., {Gottl{\"o}ber} S., 2008, \mnras, 386, 2022

\bibitem[{{Bartelmann}(1996)}]{Bartelmann96}
{Bartelmann} M., 1996, \aap, 313, 697

\bibitem[{{Battaglia} {et~al}\mbox{.}(2008){Battaglia}, {Helmi}, {Tolstoy},
  {Irwin}, {Hill}, \& {Jablonka}}]{Battaglia+08}
{Battaglia} G., {Helmi} A., {Tolstoy} E., {Irwin} M., {Hill} V., {Jablonka} P.,
  2008, \apjl, 681, L13

\bibitem[{{Beers} {et~al}\mbox{.}(1990){Beers}, {Flynn}, \& {Gebhardt}}]{BFG90}
{Beers} T.~C., {Flynn} K., {Gebhardt} K., 1990, \aj, 100, 32

\bibitem[{{Binney} \& {Mamon}(1982)}]{BM82}
{Binney} J., {Mamon} G.~A., 1982, \mnras, 200, 361

\bibitem[{{Biviano} \& {Girardi}(2003)}]{BG03}
{Biviano} A., {Girardi} M., 2003, \apj, 585, 205

\bibitem[{{Biviano} {et~al}\mbox{.}(2006){Biviano}, {Murante}, {Borgani},
  {Diaferio}, {Dolag}, \& {Girardi}}]{Biviano+06}
{Biviano} A., {Murante} G., {Borgani} S., {Diaferio} A., {Dolag} K., {Girardi}
  M., 2006, \aap, 456, 23

\bibitem[{{Biviano} \& {Poggianti}(2009)}]{BP09}
{Biviano} A., {Poggianti} B.~M., 2009, \aap, 501, 419

\bibitem[{{Borgani} {et~al}\mbox{.}(2004){Borgani}, {Murante}, {Springel},
  {Diaferio}, {Dolag}, {Moscardini}, {Tormen}, {Tornatore}, \&
  {Tozzi}}]{Borgani+04}
{Borgani} S. {et~al.}, 2004, \mnras, 348, 1078

\bibitem[{{Burkert}(1995)}]{Burkert95}
{Burkert} A., 1995, \apjl, 447, L25

\bibitem[{{Carlberg} {et~al}\mbox{.}(1997){Carlberg}, {Yee}, \&
  {Ellingson}}]{CYE97}
{Carlberg} R.~G., {Yee} H. K.~C., {Ellingson} E., 1997, \apj, 478, 462

\bibitem[{{Cuesta} {et~al}\mbox{.}(2008){Cuesta}, {Prada}, {Klypin}, \&
  {Moles}}]{CPKM08}
{Cuesta} A.~J., {Prada} F., {Klypin} A., {Moles} M., 2008, \mnras, 389, 385

\bibitem[{{Dejonghe}(1989)}]{Dejonghe89}
{Dejonghe} H., 1989, \apj, 343, 113

\bibitem[{{Dejonghe} \& {Merritt}(1992)}]{DM92}
{Dejonghe} H., {Merritt} D., 1992, \apj, 391, 531

\bibitem[{{den Hartog} \& {Katgert}(1996)}]{dHK96}
{den Hartog} R., {Katgert} P., 1996, \mnras, 279, 349

\bibitem[{{Diemand} {et~al}\mbox{.}(2004){Diemand}, {Moore}, \&
  {Stadel}}]{DMS04_vel}
{Diemand} J., {Moore} B., {Stadel} J., 2004, \mnras, 352, 535

\bibitem[{{Einasto}(1965)}]{Einasto65}
{Einasto} J., 1965, {Trudy Inst. Astroz. Alma-Ata}, 51, 87

\bibitem[{{Gerhard} {et~al}\mbox{.}(1998){Gerhard}, {Jeske}, {Saglia}, \&
  {Bender}}]{GJSB98}
{Gerhard} O., {Jeske} G., {Saglia} R.~P., {Bender} R., 1998, \mnras, 295, 197

\bibitem[{{Gerhard}(1993)}]{Gerhard93}
{Gerhard} O.~E., 1993, \mnras, 265, 213

\bibitem[{{Girardi} {et~al}\mbox{.}(1993){Girardi}, {Biviano}, {Giuricin},
  {Mardirossian}, \& {Mezzetti}}]{Girardi+93}
{Girardi} M., {Biviano} A., {Giuricin} G., {Mardirossian} F., {Mezzetti} M.,
  1993, \apj, 404, 38

\bibitem[{{Gnedin} {et~al}\mbox{.}(2004){Gnedin}, {Kravtsov}, {Klypin}, \&
  {Nagai}}]{GKKN04}
{Gnedin} O.~Y., {Kravtsov} A.~V., {Klypin} A.~A., {Nagai} D., 2004, \apj, 616,
  16

\bibitem[{{Haardt} \& {Madau}(1996)}]{HM96}
{Haardt} F., {Madau} P., 1996, \apj, 461, 20

\bibitem[{{Hansen} {et~al}\mbox{.}(2010){Hansen}, {Juncher}, \&
  {Sparre}}]{HJS10}
{Hansen} S.~H., {Juncher} D., {Sparre} M., 2010, \apjl, 718, L68

\bibitem[{{Hansen} \& {Moore}(2006)}]{HM06}
{Hansen} S.~H., {Moore} B., 2006, New Astronomy, 11, 333

\bibitem[{{Hansen} {et~al}\mbox{.}(2006){Hansen}, {Moore}, {Zemp}, \&
  {Stadel}}]{HMZS06}
{Hansen} S.~H., {Moore} B., {Zemp} M., {Stadel} J., 2006, \jcap, 1, 14

\bibitem[{{Hernquist}(1990)}]{Hernquist90}
{Hernquist} L., 1990, \apj, 356, 359

\bibitem[{{Katgert} {et~al}\mbox{.}(2004){Katgert}, {Biviano}, \&
  {Mazure}}]{KBM04}
{Katgert} P., {Biviano} A., {Mazure} A., 2004, \apj, 600, 657

\bibitem[{{Kronawitter} {et~al}\mbox{.}(2000){Kronawitter}, {Saglia},
  {Gerhard}, \& {Bender}}]{KSGB00}
{Kronawitter} A., {Saglia} R.~P., {Gerhard} O., {Bender} R., 2000, \aaps, 144,
  53

\bibitem[{{Lemze} {et~al}\mbox{.}(2012){Lemze}, {Wagner}, {Rephaeli}, {Sadeh},
  {Norman}, {Barkana}, {Broadhurst}, {Ford}, \& {Postman}}]{Lemze+12}
{Lemze} D. {et~al.}, 2012, \apj, 752, 141

\bibitem[{{Lewis} \& {Bridle}(2002)}]{LB02}
{Lewis} A., {Bridle} S., 2002, \prd, 66, 103511

\bibitem[{{{\L}okas}(2002)}]{Lokas02}
{{\L}okas} E.~L., 2002, \mnras, 333, 697

\bibitem[{{{\L}okas} \& {Mamon}(2001)}]{LM01}
{{\L}okas} E.~L., {Mamon} G.~A., 2001, \mnras, 321, 155

\bibitem[{{{\L}okas} \& {Mamon}(2003)}]{LM03}
{{\L}okas} E.~L., {Mamon} G.~A., 2003, \mnras, 343, 401

\bibitem[{{Ludlow} {et~al}\mbox{.}(2009){Ludlow}, {Navarro}, {Springel},
  {Jenkins}, {Frenk}, \& {Helmi}}]{Ludlow+09}
{Ludlow} A.~D., {Navarro} J.~F., {Springel} V., {Jenkins} A., {Frenk} C.~S.,
  {Helmi} A., 2009, \apj, 692, 931

\bibitem[{{Macci{\`o}} {et~al}\mbox{.}(2008){Macci{\`o}}, {Dutton}, \& {van den
  Bosch}}]{MDvdB08}
{Macci{\`o}} A.~V., {Dutton} A.~A., {van den Bosch} F.~C., 2008, \mnras, 391,
  1940

\bibitem[{{Mamon} {et~al}\mbox{.}(2010){Mamon}, {Biviano}, \&
  {Murante}}]{MBM10}
{Mamon} G.~A., {Biviano} A., {Murante} G., 2010, \aap, 520, A30

\bibitem[{{Mamon} \& {Bou{\'e}}(2010)}]{MB10}
{Mamon} G.~A., {Bou{\'e}} G., 2010, \mnras, 401, 2433

\bibitem[{{Mamon} \& {{\L}okas}(2005)}]{ML05b}
{Mamon} G.~A., {{\L}okas} E.~L., 2005, \mnras, 363, 705

\bibitem[{{Mauduit} \& {Mamon}(2007)}]{MM07}
{Mauduit} J.-C., {Mamon} G.~A., 2007, \aap, 475, 169

\bibitem[{{Merritt}(1985)}]{Merritt85_df}
{Merritt} D., 1985, \mnras, 214, 25P

\bibitem[{{Merritt}(1987)}]{Merritt87}
{Merritt} D., 1987, \apj, 313, 121

\bibitem[{{Merritt} \& {Saha}(1993)}]{MS93}
{Merritt} D., {Saha} P., 1993, \apj, 409, 75

\bibitem[{{Navarro} {et~al}\mbox{.}(1996){Navarro}, {Frenk}, \&
  {White}}]{NFW96}
{Navarro} J.~F., {Frenk} C.~S., {White} S. D.~M., 1996, \apj, 462, 563

\bibitem[{{Navarro} {et~al}\mbox{.}(2004){Navarro}, {Hayashi}, {Power},
  {Jenkins}, {Frenk}, {White}, {Springel}, {Stadel}, \& {Quinn}}]{Navarro+04}
{Navarro} J.~F. {et~al.}, 2004, \mnras, 349, 1039

\bibitem[{{Newman} {et~al}\mbox{.}(2009){Newman}, {Treu}, {Ellis}, {Sand},
  {Richard}, {Marshall}, {Capak}, \& {Miyazaki}}]{Newman+09}
{Newman} A.~B., {Treu} T., {Ellis} R.~S., {Sand} D.~J., {Richard} J.,
  {Marshall} P.~J., {Capak} P., {Miyazaki} S., 2009, \apj, 706, 1078

\bibitem[{{Osipkov}(1979)}]{Osipkov79}
{Osipkov} L.~P., 1979, Soviet Astronomy Letters, 5, 42

\bibitem[{{Powell}(2006)}]{Powell06}
{Powell} M.~J.~D., 2006, in Nonconvex Optimization and Its Applications, {Di
  Pillo} G., {Roma} M., eds., Vol.~83, Spinger, pp. 255--297

\bibitem[{{Richardson} \& {Fairbairn}(2012)}]{RF12}
{Richardson} T., {Fairbairn} M., 2012, \mnras, submitted, arXiv:1207.1709

\bibitem[{{Richstone} \& {Tremaine}(1984)}]{RT84}
{Richstone} D.~O., {Tremaine} S., 1984, \apj, 286, 27

\bibitem[{{Sanchis} {et~al}\mbox{.}(2004){Sanchis}, {{\L}okas}, \&
  {Mamon}}]{SLM04}
{Sanchis} T., {{\L}okas} E.~L., {Mamon} G.~A., 2004, \mnras, 347, 1198

\bibitem[{{Schwarzschild}(1979)}]{Schwarzschild79}
{Schwarzschild} M., 1979, \apj, 232, 236

\bibitem[{{Sparre} \& {Hansen}(2012)}]{SH12}
{Sparre} M., {Hansen} S.~H., 2012, \jcap, 10, 49

\bibitem[{{Springel} \& {Hernquist}(2003)}]{SH03}
{Springel} V., {Hernquist} L., 2003, \mnras, 339, 289

\bibitem[{{Springel} {et~al}\mbox{.}(2005){Springel}, {White}, {Jenkins},
  {Frenk}, {Yoshida}, {Gao}, {Navarro}, {Thacker}, {Croton}, {Helly},
  {Peacock}, {Cole}, {Thomas}, {Couchman}, {Evrard}, {Colberg}, \&
  {Pearce}}]{Springel+05}
{Springel} V. {et~al.}, 2005, \nat, 435, 629

\bibitem[{{Strigari} {et~al}\mbox{.}(2008){Strigari}, {Bullock}, {Kaplinghat},
  {Simon}, {Geha}, {Willman}, \& {Walker}}]{Strigari+08}
{Strigari} L.~E., {Bullock} J.~S., {Kaplinghat} M., {Simon} J.~D., {Geha} M.,
  {Willman} B., {Walker} M.~G., 2008, \nat, 454, 1096

\bibitem[{{Syer} \& {Tremaine}(1996)}]{ST96}
{Syer} D., {Tremaine} S., 1996, \mnras, 282, 223

\bibitem[{{Tiret} {et~al}\mbox{.}(2007){Tiret}, {Combes}, {Angus}, {Famaey}, \&
  {Zhao}}]{Tiret+07}
{Tiret} O., {Combes} F., {Angus} G.~W., {Famaey} B., {Zhao} H.~S., 2007, \aap,
  476, L1

\bibitem[{{Tsallis}(1988)}]{Tsallis88}
{Tsallis} C., 1988, Journal of Statistical Physics, 52, 479

\bibitem[{{van der Marel}(1994)}]{vanderMarel94}
{van der Marel} R.~P., 1994, \mnras, 270, 271

\bibitem[{{van der Marel} \& {Franx}(1993)}]{vdMF93}
{van der Marel} R.~P., {Franx} M., 1993, \apj, 407, 525

\bibitem[{{Wainer} \& {Thissen}(1976)}]{WT76}
{Wainer} H., {Thissen} D., 1976, Psychometrica, 41, 9

\bibitem[{{Walker} {et~al}\mbox{.}(2009){Walker}, {Mateo}, {Olszewski},
  {Pe{\~n}arrubia}, {Evans}, \& {Gilmore}}]{Walker+09}
{Walker} M.~G., {Mateo} M., {Olszewski} E.~W., {Pe{\~n}arrubia} J., {Evans}
  N.~W., {Gilmore} G., 2009, \apj, 704, 1274

\bibitem[{{Wojtak} {et~al}\mbox{.}(2005){Wojtak}, {{\L}okas}, {Gottl{\"o}ber},
  \& {Mamon}}]{WLGM05}
{Wojtak} R., {{\L}okas} E.~L., {Gottl{\"o}ber} S., {Mamon} G.~A., 2005, \mnras,
  361, L1

\bibitem[{{Wojtak} {et~al}\mbox{.}(2009){Wojtak}, {{\L}okas}, {Mamon}, \&
  {Gottl{\"o}ber}}]{WLMG09}
{Wojtak} R., {{\L}okas} E.~L., {Mamon} G.~A., {Gottl{\"o}ber} S., 2009, \mnras,
  399, 812

\bibitem[{{Wojtak} {et~al}\mbox{.}(2008){Wojtak}, {{\L}okas}, {Mamon},
  {Gottl\"ober}, {Klypin}, \& {Hoffman}}]{Wojtak+08}
{Wojtak} R., {{\L}okas} E.~L., {Mamon} G.~A., {Gottl\"ober} S., {Klypin} A.,
  {Hoffman} Y., 2008, \mnras, 388, 815

\bibitem[{{Wojtak} {et~al}\mbox{.}(2007){Wojtak}, {{\L}okas}, {Mamon},
  {Gottl{\"o}ber}, {Prada}, \& {Moles}}]{Wojtak+07}
{Wojtak} R., {{\L}okas} E.~L., {Mamon} G.~A., {Gottl{\"o}ber} S., {Prada} F.,
  {Moles} M., 2007, \aap, 466, 437

\bibitem[{{Wojtak} \& {Mamon}(2013)}]{WM13}
{Wojtak} R., {Mamon} G.~A., 2013, \mnras, 428, 2407

\bibitem[{{Wolf} {et~al}\mbox{.}(2010){Wolf}, {Martinez}, {Bullock},
  {Kaplinghat}, {Geha}, {Mu{\~n}oz}, {Simon}, \& {Avedo}}]{Wolf+10}
{Wolf} J., {Martinez} G.~D., {Bullock} J.~S., {Kaplinghat} M., {Geha} M.,
  {Mu{\~n}oz} R.~R., {Simon} J.~D., {Avedo} F.~F., 2010, \mnras, 406, 1220

\bibitem[{{Zabludoff} {et~al}\mbox{.}(1993){Zabludoff}, {Franx}, \&
  {Geller}}]{ZFG93}
{Zabludoff} A.~I., {Franx} M., {Geller} M.~J., 1993, \apj, 419, 47

\bibitem[{{Zhao}(1996)}]{Zhao96}
{Zhao} H., 1996, \mnras, 278, 488

\end{thebibliography}

\onecolumn
\appendix

\section{Kernels for determining the radial velocity dispersion profile from
  the Jeans equation}

\label{kernels}
The radial squared velocity dispersion profile of equation~(\ref{sigmar2})
can be  written \citep{vanderMarel94,ML05b}
\begin{equation}
\sigma_r^2(r) = {1\over K(r)\,\nu(r)}\,\int_r^\infty K(s)\,\nu(s){G
  M(s)\over s^2}\,\d s \ ,
\label{sigmar2K}
\end{equation}
where $K(r)$ is the solution to ${\rm d}\ln K/\d \ln r=2\,\beta(r)$, i.e.
\begin{equation}
{K(r)\over K(s)} = \exp\left [2\int_r^s \!\! \beta(t) {\d t\over t} \right ]
\ .
\label{KoverK}
\end{equation}

For simple anisotropy models (not all are used in this work), 
\begin{equation}
K(r) = 
\left \{
\begin{array}{ll}
\displaystyle
r^{2\beta}
& \beta={\rm cst} ,\\
& \\
\displaystyle
r^2+r_\beta^2
& \displaystyle
\beta \equiv \beta_{\rm OM} = {r^2\over
  r^2+r_\beta^2} ,\\
& \\
\displaystyle
r+r_\beta &
\displaystyle
\beta \equiv \beta_{\rm ML} = {1\over2}\,{r\over
  r+r_\beta} ,\\
& \\
\displaystyle
(r+r_\beta)^{2\,\beta_\infty} &
\displaystyle
\beta \equiv \beta_{\rm T} 
= \beta_\infty\, {r\over r\!+\!r_\beta}
, \\
\\
\displaystyle
r^{2\,\beta_0}\,(r+r_\beta)^{2\,(\beta_\infty-\beta_0)} &
\displaystyle
\beta \equiv \beta_T^{\rm gen} = \beta_0 + \left (\beta_\infty-\beta_0\right)\,{r\over r+r_\beta}
, \\
\\
\displaystyle
\exp \left [6\,\left ({r\over r_\beta}\right)^{1/3}\right] 
& \beta \equiv \beta_{\rm DMS} \ (r < r_\beta) , \\
& \\
\displaystyle
{\rm e}^6\,\left ({r\over r_\beta}\right)^2\,
& \beta \equiv \beta_{\rm DMS} \ (r \geq r_\beta) , \\
& \\
\displaystyle
a + b\,\left ({\gamma_0\,r_{\overline\gamma}+\gamma_\infty\,r \over r+r_{\overline\gamma}}\right )
& \displaystyle \beta = a + b \,{{\rm d} \ln \nu \over {\rm d}\ln r} \ ,
\qquad
(\hbox{Hansen \& Moore / Zhao}) \\
& \\
\displaystyle
r^{2a}\,\exp \left (-4 \,b\,m\, r^{1/m}\right)
& \displaystyle \beta = a + b \,{{\rm d} \ln \nu \over {\rm d}\ln r}\ ,
\qquad (\hbox{Hansen \& Moore / Einasto})
\end{array}
\right.
\label{exp2intbetaovert}
\end{equation}
where
$\beta_{\rm OM}$ is the Osipkov-Merritt anisotropy model
\citep{Osipkov79,Merritt85_df}, 
$\beta_{\rm ML}$ is the Mamon-{\L}okas  anisotropy model \citep{ML05b},
$\beta_{\rm T}^{\rm gen}$ is a generalization of the Mamon-{\L}okas model by
\cite{Tiret+07},  
$\beta_{\rm DMS} = {\rm Min}\left [1,(r/r_\beta)^{1/3}\right]$ is another model that fits
well the anisotropy of $\Lambda$CDM haloes with $r_\beta=r_v$ \citep*{DMS04_vel},
while the last two rows describe the anisotropy that varies linearly with the
slope of the density profile \citep{HM06}, since the kernel depends on the
precise form of the density profile.
The first of these last two rows gives the kernel for
\citeauthor{Zhao96}'s 
(\citeyear{Zhao96}) general density profile
\begin{equation}
\rho(r) \propto \left ({r\over r_{\overline\gamma}}\right)^{\gamma_0} \,\left ({r\over
  r_{\overline\gamma}} + 1\right)^{\gamma_\infty-\gamma_0}
\ ,
\end{equation}
where $\gamma_0$ and $\gamma_\infty$ are the inner and outer logarithmic
slopes of the density profile  
(i.e. $\gamma_0 = -1$ and $\gamma_\infty=-3$ for the NFW
model), and $r_{\overline\gamma} = -[(2+\gamma_\infty)/(2+\gamma_0)]\,r_{-2}$
is the radius of slope $\overline\gamma=(\gamma_0+\gamma_\infty)/2$, while
$r_{-2}$ is the radius of slope $-2$.
The last row is for the \cite{Einasto65} density profile
\begin{equation}
\rho(r) \propto \exp \left [-2\,m\left ({r\over r_{-2}}\right)^{1/m} \right]
\ ,
\end{equation}
which, for $m \approx 6$, is an excellent fit to the density profiles of
simulated $\Lambda$CDM 
haloes \citep{Navarro+04}, while $m=5$ is an excellent fit to the dark matter
density profiles of haloes in hydrodynamical $\Lambda$CDM cosmological
simulations (MBM10).

\section{Interloper removal}
\label{clean}
This appendix describes how we remove obvious (high absolute line-of-sight
velocity) interlopers.

We analyze the DM particles in the $z=0$ output of the
hydrodynamical 
cosmological simulation of \cite{Borgani+04}, to which we
  added the Hubble flow, placed an observer at $90 \, h^{-1} \, \rm Mpc$ from
  each of the 105 most massive haloes (including the most irregular ones), in
  each of the three cartesian directions. 
We limited these $3\times105=315$ haloes to projected radii
within $R_{\rm max} = X_{\rm max} \,r_{200}$ from the barycentre of the
Friends-of-Friends halo identified in real space, where $X_{\rm
    max} = 0.7$, 1 or 1.4, and within line-of-sight velocities within 4 times the
  true circular velocity at the $r_{200}$ from the true velocity centre (see
  {MBM10} for details). The choice for the three values of $X_{\rm max}$ is
  to consider cases where the observers guess incorrectly the virial radius
  when they select their galaxies.
We only used the 274 haloes out of 315 that had at least 500 particles
with $R<r_{200}$.\footnote{We used an output of the simulation with 1~
  particle 
out of 55 chosen at random.}

We chose $N$ particles at random in the projected phase space, so that we ended up with
  approximately $n=500$ particles with $R<r_{200}$, i.e. $N=500$ for $X_{\rm
    max}=1$ or $N=500\,M_{\rm p}^{\rm NFW}(X_{\rm max} r_{200})/M_{\rm p}^{\rm
    NFW}(r_{200}) = 386$ and 620 for $X_{\rm max}=0.7$ and 1.4, respectively, 
where we assumed a concentration
  $c=r_{200}/r_{-2}=4$, typical of the haloes of our simulation (MBM10).
The projected mass of the NFW model is $M_{\rm p}(R) = M(r_{-2})\,\widetilde
M_{\rm p}(R/r_{-2})$, where $r_{-2}$ is the radius where the logarithmic  of
the density profile is $-2$ and the dimensionless projected mass is
 (\citealp{LM01}, first derived by
\citealp{Bartelmann96} in a slightly longer and more computer intensive form)
\begin{equation}
\widetilde M_{\rm p}(X) = {1\over \ln 2 -1/2}\,\left\{
\begin{array}{ll}
\cosh^{-1}(1/X)/\sqrt{1-X^2}+\ln(X/2) & \hbox{\qquad for } X<1 \ ,\\
 1-\ln2 & \hbox{\qquad for } X=1\ ,\\
\cos^{-1}(1/X)/\sqrt{X^2-1}+\ln(X/2) & \hbox{\qquad for } X>1 \ .
\end{array}
\right.
\label{mproj}
\end{equation}

Our algorithm for interloper rejection, which is Bayesian as it assumes
what we know about $\Lambda$CDM haloes, namely that they approximately follow NFW
density profiles with increasingly radial orbits, goes as follows.

\renewcommand{\theenumi}{\roman{enumi})}
\begin{enumerate}
\itemsep 0.5\baselineskip
\item \label{gapper}
On first pass, we apply the \emph{gapper} (or \emph{weighted gap}) 
technique in order to split multimodal distributions of the
distribution of velocities $v_i$, and identify the peak
in velocity space closest to the input mean velocity of the
halo, which we assume to know.
After sorting the velocity offsets, we compute weighted gaps
\begin{equation}
{\cal G}_i = [i\,(n-i)\, \left(v_{i+1}-v_i\right)]^{1/2}
\end{equation}
for $1 \leq i < n$.
We then check if the largest value of the dimensionless gap 
${\cal G}_i/{\rm MidMean}({\cal G})$ is
greater than some threshold, commonly called $C$
(MidMean is the arithmetic mean within the quartiles of the distribution).
Among all the subsamples separated by a dimensionless gap $\geq C$,
we keep only that one closest to the input mean velocity of the
halo.
The gapper technique \citep{WT76} was first used in astronomy in the ROSTAT
package\footnote{By T. Beers, see
  http://www.pa.msu.edu/~ftp/pub/beers/posts/rostat/}, which recommends
$C=2.25$ 
and was first applied to detect multimodal
populations in clusters of galaxies by \cite{Girardi+93}, who used
$C=4$. This is also the value we use here.

\item \label{sigmad} Using the velocities of the particles in the
  subsample identified with the gapper technique, we compute, in a
  first step, the global velocity dispersion of our selected
  particles, using the robust median absolute deviation (MAD), with
  $\sigma_{\rm v,MAD} \simeq {\rm MAD}(v)/0.6745$ where ${\rm
    MAD}={\rm Median}|v-{\rm Median}(v)|$ (e.g. \citealp{BFG90}).

\item \label{vvirest}
We then estimate the virial velocity , $v_{\rm v}^{\rm est} = v_{\rm
  circ}(r_{\rm v}^{\rm est})$
from the relation
\begin{eqnarray}
\left ({\sigma_{\rm ap} (r_{\rm v}) \over v_{\rm v}} \right)^2
&=& 
{[1/M_{\rm p}(r_{\rm v})]\,\int_0^{r_{\rm v}} 2 \pi
R\,\Sigma(R)\,\sigma_z^2(R)\,{\rm d}R 
\over
G \,M(r_{\rm v})/r_{\rm v}}
\nonumber \\
&=& 4{\,\pi\,r_{\rm v}\over M(r_{\rm v})\,M_{\rm p}(r_{\rm
    vir})}\,\int_0^{r_{\rm v}} R\,{\rm d}R\,\int_R^\infty K_\beta\left
({r\over R},{r_\beta \over R} \right)\,\rho(r)\,M(r)\,{{\rm d}r\over r}
\label{sigrat1}\\
&=&
{c\over \widetilde M(c) \widetilde M_{\rm p}(c)}\,
\int_0^c X\,{\rm d}X\,\int_X^\infty 
K_\beta \left ({x\over X},{x_\beta \over X} \right)
\,\widetilde \rho(x)\,\widetilde M(x)\,{{\rm d}x\over x} \ ,
\label{sigrat2}
\end{eqnarray}
where 
$\widetilde \rho(x) = \rho(r_{-2}\,x)\,/\,[ M(r_{-2})/(4\pi r_{-2}^3)]$,
$\widetilde M(x) = M(r_{-2}\,x) / M(r_{\rm -2})$,
and
$\widetilde M_{\rm p}(x) = M_{\rm p}(r_{-2}\,x) / M(r_{\rm -2})$ are the
dimensionless radial profiles of density, mass and projected mass, respectively.
Equation~(\ref{sigrat1})  was derived using the relation \citep{ML05b}
\begin{equation}
\Sigma(R)\,\sigma_z^2(R) = 2\,G\,\int_R^\infty K_\beta\left
({r\over R},{r_\beta \over R} \right)\,\rho(r)\,M(r)\,{{\rm d}r\over r}
\end{equation}
with
$K_\beta$ a kernel that has been derived by \citeauthor{ML05b} for simple
anisotropy profiles.
In equation~(\ref{sigrat2}),
we 
assume 
ML anisotropy (eq.~[\ref{e:ML}]) 
with $r_\beta = r_{-2}$
as found in a cosmological simulation by MBM10. We then derive $v_{\rm
  v}=v_{\rm circ}(r_{\rm v})$ from equation~(\ref{sigrat2}).
We adopt the NFW model, with
dimensionless mass density $\widetilde \rho(x) = x^{-1}
(x+1)^{-2}/(\ln2-1/2)$,
dimensionless mass $\widetilde M(x) = [\ln(x+1)-x/(x+1)]/(\ln2-1/2)$
and dimensionless projected mass density  $M_{\rm p}(X)$ given in
equation~(\ref{mproj}).
With the \citeauthor{ML05b} anisotropy profile, and $r_\beta = r_{-2}$,
the aperture velocity dispersion is well approximated (to better than
0.1\% relative accuracy in the range $1 \leq c < 32$) by
\begin{equation}
{\sigma_{\rm ap} (r_{\rm v})\over v_{\rm v}} \simeq 
{\rm dex} \left [\sum_{i=0}^3 a_i (\log c)^i \right] \ ,
\label{sigap}
\end{equation}
where $a_0 = -0.1197$, $a_1=-0.2176$, $a_2=0.2082$, and $a_3=-0.03087$.
The ratio
$\sigma_{\rm ap}(r_{\rm v})/v_{\rm v}$ reaches a minimum of 0.65 at
$c=4$, with values of 0.69 at $c=1.8$ and 10.
Had we assumed isotropy instead, $\sigma_{\rm ap}(r_{\rm v})/v_{\rm v}$
would have been 3\% lower \citep{MM07}.
In this first pass, we assume $c=4$, while in subsequent passes we use the
relation obtained in $\Lambda$CDM haloes by \cite{MDvdB08}: $c =
6.76\,(h\,M/10^{12}\rm M_\odot)^{-0.098}$.

\item \label{rvirest} 
We deduce an estimated virial radius
$r_{\rm v}^{\rm est} = \sqrt{\Delta/2} H_0 \,v_{\rm v}^{\rm est}$, with
  overdensity $\Delta=200$. We use $H_0 = 100 \,\rm km \,s^{-1} \, Mpc^{-1}$ to
  conform with the positional units of the simulation (and which we had
  thus assumed for the Hubble flow).

\item \label{siglosall}
Next, we compute LOS velocity dispersion predicted for the NFW model with
anisotropy profile of equation~(\ref{e:ML}) at the projected radius of
  every particle. For this, we use the approximation 
\begin{equation}
{\sigma_z^{\rm NFW}(R) \over \sqrt{G M(r_{-2})/r_{-2}}}
\simeq
{\rm dex} \left\{\sum_{i=0}^7 b_i \left[\log \left ({R\over
    r_{-2}}\right)\right] ^i \right\}
\ ,
\end{equation}
with the coefficients $b_i$ given in Table~2 (for the NFW column) of
MBM10
and
\begin{equation}
{G M(r_{-2})/r_{-2}\over v_{\rm v}^2} = {(\ln 2 -1/2)\,c\over
  \ln(c+1)-c/(c+1)}
\end{equation}
(e.g., eq.~[22] of MBM10). 

\item \label{vfilter}
We filter the particles to have velocities within $\kappa$ times
  the local LOS velocity dispersion from the global median velocity.
We adopt $\kappa=2.7$, which best preserves the LOS velocity dispersion profile
(MBM10).

\item \label{sigv}
We compute the global velocity dispersion of our velocity-filtered
  sample, this time using the standard unbiased standard deviation.

\item \label{iter}
We iterate, checking (except after the first pass) that the number of
particles has changed or that $r_{\rm v}^{\rm est}$
had changed by more than 0.1\%, by returning to step~\ref{vvirest}.

\item On convergence, we select all particles within $2.7\,\sigma_z(R)$
  from the median (except for those filtered out by the gapper technique).
\end{enumerate}

The novelty of this algorithm is that 1) it uses a guess of the (local) LOS
velocity dispersion, and 2) it uses a cut at $2.7\,\sigma$ instead of 3 to
optimally recover $\sigma_z(R)$ (MBM10).

\end{document}